\documentclass[sigconf,balance=false]{acmart}
\usepackage{popets}
\usepackage{spverbatim}
\usepackage{booktabs}
\usepackage{subcaption}
\usepackage{array}
\usepackage{todonotes}

\setcopyright{popets}
\copyrightyear{YYYY}

\acmYear{YYYY}
\acmVolume{YYYY}
\acmNumber{X}
\acmDOI{XXXXXXX.XXXXXXX}
\acmISBN{}
\acmConference{Proceedings on Privacy Enhancing Technologies}
\begin{document}

\title[Automatic Privacy Analysis and Legal Compliance of TCF-based Android Applications]{The TCF doesn't really A(A)ID -- Automatic Privacy Analysis and Legal Compliance of TCF-based Android Apps}


\author{Victor Morel}
\affiliation{%
  \institution{Chalmers University of Technology and University of Gothenburg}
  \city{Gothenburg}
  \country{Sweden}
  \state{}}
\email{morelv@chalmers.se}

\author{Cristiana Santos}
\affiliation{%
  \institution{Utrecht University}
  \city{Utrecht}
  \country{The Netherlands}
  \state{}}
\email{c.teixeirasantos@uu.nl}

\author{Pontus Carlsson}
\affiliation{%
  \institution{Chalmers University of Technology and University of Gothenburg}
  \city{Gothenburg}
  \country{Sweden}
  \state{}}
\email{pontus@carlsson4.se}

\author{Joel Ahlinder}
\affiliation{%
  \institution{Chalmers University of Technology and University of Gothenburg}
  \city{Gothenburg}
  \country{Sweden}
  \state{}}
\email{joelahlinder@hotmail.com}

\author{Romaric Duvignau}
\affiliation{%
  \institution{Chalmers University of Technology and University of Gothenburg}
  \city{Gothenburg}
  \country{Sweden}
  \state{}}
\email{duvignau@chalmers.se}


\renewcommand{\shortauthors}{Morel et al.}

\begin{abstract}
The Transparency and Consent Framework (TCF), developed by the Interactive Advertising Bureau (IAB) Europe, provides a \textit{de facto} standard for requesting, recording, and managing user consent from European end-users.
Its goal is to help organizations comply with the General Data Protection Regulation (GDPR) and the ePrivacy Directive (ePD).
  %
This framework has previously been found to infringe European data protection law and has subsequently been regularly updated.
Previous research on the TCF focused exclusively on web contexts, with no attention given to its implementation in mobile apps. 
No work has systematically studied the compliance implications of the TCF on Android apps.
To address this gap, we investigate the prevalence of the TCF in popular Android apps from the Play Store, and assess whether these apps respect users' consent banner choices.
The TCF introduced minor changes in its new version (v.2.3) on March 1st, 2026, aimed at reducing ambiguity for vendors; these changes do not impact our results.

  
We scraped and downloaded 4482 of the most popular Google Play Store apps on an emulated Android device.
We automatically identified that 576 (12.85\%) of the 4482  downloadable apps  implemented the TCF, 
 and we detected potential legal violations within this subset.
  %
By automatically interacting with consent banners, we observed that in 15 (2.6\%) of these apps, 
  users’ choices are stored only when consent is granted. Users who refuse consent are shown the consent banner again each time they launch the app.
  %
We analyzed the apps' traffic in two different stages, passive (post interaction with the banner) and active (during banner interaction and post user choices).
  Network analysis conducted during  the passive stage reveals  that 66.2\% of the analyzed TCF-based apps share  personal data through  the Google Advertising ID (AAID) without consent -- the lawful basis for such processing.
  Furthermore, 55.3\% of apps analyzed during the active stage share AAID before users interact with the apps' consent banners, violating the prior consent requirement.  
  We further expose concerns regarding Google as the dominant Consent Management Provider (CMP) in our dataset (89.76\%), structurally accommodating potential legal violations. 
   Our results suggest that mobile implementations of the TCF are prone to significant non-compliance practices, raising  concerns about its effectiveness in helping organizations adhere to legal requirements.
\end{abstract}



\maketitle


\section{Introduction}


The majority of Web traffic now originates from mobile devices~\cite{similarweb}, and in the past five years, Android has remained the most widely used mobile operating system (72.55\% in October 2025)~\cite{mobile-os-marketshare}. 
It has become increasingly common to rely on mobile apps -- generally distributed through the Google Play Store --, to communicate and interact with devices and services, a phenomenon known as \textit{appification}~\cite{case_appification_2019}.
However, the widespread use of these apps is accompanied by a large-scale tracking impacting users' privacy~\cite{ap_news_google_2025}.
%
In the European Union (EU), personal data collection and processing is strictly regulated.
The General Data Protection Regulation (GDPR)~\cite{GDPR2018} and the ePrivacy Directive (ePD)~\cite{ePD-09} specify legal requirements for data controllers (herein mobile app publishers\footnote{The French DPA (CNIL) qualifies app publishers as data controllers, and note that publishers and developers often merge in practice, although external developers can be considered a different contractual entity~\cite[p17,p7] {cnil_recommendation_2025}.} and website owners) to track end-users's personal data. 
In particular, data protection law mandates that controllers identify 
a valid legal basis for each purpose of processing 
under Article 6 of the GDPR.
%
Regulatory  and enforcement decisions have consistently affirmed that consent  is  the  \textit{only}  lawful  basis  for  online  behavioral advertising, instead of legitimate interest~\cite{CJEU2023_Bundeskartellamt, WP29_2010_OnlineBehaviouralAdvertising}.  

To help organizations request user consent according to the legal requirements,
the Interactive Advertising Bureau (IAB) Europe --  the leading industry association for the digital marketing and advertising ecosystem in Europe -- 
introduced in April 2018 the Transparency and Consent Framework (TCF)~\cite{TCF}. 
The TCF is widely seen as the leading industry standard for managing online consent~\cite{waem_eu_2025}, although academic literature demonstrates instances of its non-compliance practices on the Web 
~\cite{matte2020cookie,bollinger2022automating,MatteSantosBielova2020,kancherla_johnny_2025}, but even less so on mobile contexts~\cite{koch2023ok}.
%
Although the TCF purports to help organizations meet EU data protection law obligations, its deployment does not guarantee compliance. 
Prior analysis has shown that websites relying on earlier  TCF versions (\textit{i.e.}, v.1.1) violated data protection law despite formally implementing the framework~\cite{matte2020cookie, morel2023legitimate, Veale2022ImpossibleAsks}. 
Specifically, the Belgian Data Protection Authority (APD) ruled in 2022~\cite{BelgianDPA2022_Decision21_2022} that the TCF was non-compliant with the GDPR as it allowed the use of “legitimate interest” for targeted advertising -- a legal basis deemed incompatible with the intrusive nature of large-scale behavioral tracking and profiling. 
The APD therefore required IAB Europe to remove “legitimate interest” as a valid ground for processing data for advertising purposes. 
Furthermore, the ruling concluded that the framework’s consent mechanism failed to meet the GDPR consent requirements. The EU Court of Justice  held that the TC String, used to record users’ consent preferences, can constitute personal data and it also found that IAB Europe may act as a joint controller with TCF participants even if it does not directly access the data~\cite{CJEU_IAB_TCF_2024}. 

The framework has undergone several revisions since these rulings. 
The update (v.2.2) introduced in May 2023 required all TCF participants to implement the new version by November 2023. 
A major change introduced in this update was the removal of “legitimate interest” as a legal basis for processing personal data to create personalized ads and content~\cite{TCF-v.2.2}. While this reform was intended to be legally compliant, it remains unclear whether the framework is consistently implemented and compliant.
V.2.3 superseded v.2.2 on March 1st 2026, without incidence on our results.




In practice, respect of GDPR requirements appears to have a long way to go on mobile.
Previous research on the GDPR compliance of Android apps found 
that 28\% of 86000 apps potentially violated the GDPR by sending users' personal data to ad providers before  users' consent was obtained~\cite{nguyen2021share}. 
Furthermore, TCF-based websites have frequently been shown to contain potential privacy violations~\cite{matte2020cookie}, suggesting that similar issues may also occur in  TCF-based Android apps.
%
Despite these findings, research on TCF implementations in Android apps remains limited, and no prior studies have examined their use of legitimate interest as a legal basis for data processing. 
To address this gap, we aim to investigate TCF-based Android apps by answering the following research questions:


\begin{enumerate}
    \item[\textbf{RQ1}] How prevalent are TCF-based apps in the Google Play Store, and what are their characteristics (\textit{e.g.}, categories, country of origin, etc.)? 
    \item[\textbf{RQ2}] Do TCF-based apps respect users' consent choices and the TCF's requirements introduced by the v.2.2 not to use legitimate interest for certain purposes?
    \item[\textbf{RQ3}] Do TCF-based Android apps respect users' choices regarding personal data collection at runtime? 
\end{enumerate}

To address these questions, we collaborated with a legal scholar in EU Data Protection law
and we make the following contributions:

\begin{enumerate}
    \item We measure the prevalence of  TCF in popular Android apps, including the first fine-grain quantification per category, and find a significant increase in its adoption, rising from 6.6\%~ reported in 2023~\cite{koch2023ok} to 12.85\% in our dataset.   

    \item We design a methodology to semi-automatically interact with consent banners deployed by  major Consent Management Providers (CMPs) in TCF-based Android apps, combining three approaches capturing different choices to data collection: ``disagree to all'' purposes, ``legitimate interest purposes only'', and ``consent to all''. 
    \item We perform the first systematic assessment of personal data processing in TCF-based Android apps through traffic analysis. 
    Our results indicate that 15 apps stored consent choices only when provided with ``consent to all'' data processing, and that most apps (66.2\%), with gaming apps as the top violators, share AAID to ad servers against users' choices. We perform a legal analysis to assess compliance with GDPR and ePD requirements, which exposes several potential legal violations. These results suggest a systemic breakdown in the ability of the TCF to facilitate legal compliance, questioning its effectiveness as a privacy-governance mechanism. 
    \item We expose concerns regarding Google as the dominant CMP in our dataset (89.76\%), structurally accommodating potential legal violations, and offering bad defaults (pre-selected legitimate interest-based purposes). 
\end{enumerate}


\section{Background knowledge}
\label{sec:background}
This section introduces background related to GDPR and the ePD, AAID, and TCF.

\subsection{EU Data Protection Law}
\label{subsec:gdpr-epd}
The GDPR came into force in 2018, and applies to any organization worldwide that processes personal data from EU users~\cite{EDPB-4-07}, called data controllers in this context.
Data controllers (app publishers in our context) ``determine the purposes and means of the processing of personal data'', they bear legal responsibility for processing.
The GDPR imposes obligations on data controllers, paired with hefty fines for non-compliance.
The ePrivacy Directive (ePD) provides~\emph{supplementary} rules to the GDPR, in particular for the use of  tracking technologies. 
Whenever any information (including cookies and other tracking technologies) is stored and read from the user's device, the ePD~\cite[Art. 5(3)]{ePD-09} requires  
organizations to request user \textit{consent} 
for certain \emph{purposes}, such as advertising~\cite{EDPB-2-13,EDPB-3-13} to process data.
The only way to assess with certainty whether consent is required is to analyze the \emph{purpose} of each tracker on a given website (or app in our context)~\cite{EDPB-4-12,Foua-etal-20-IWPE, santos2021cookie}.

\noindent \textbf{Personal data.}
Personal data is
\textit{``any information relating to an identified or identifiable natural person ('data subject'). 
An identifiable natural person is one who can be identified, directly or indirectly, in particular by reference to an identifier such as a name, an identification number, location data, an online identifier or to one or more factors specific to the physical, physiological, genetic, mental, economic, cultural or social identity of that natural person''}~\cite[Art. 4(1)]{GDPR2018}.
In order to determine whether a person is \textit{identifiable}, account should be taken of all the means likely to be reasonably used by the  data controller (herein the app publisher) to identify that person (Recital 26 GDPR).
Account should be taken by this actor’s specific situation -- \textit{i.e.}, whether it has reasonably likely means to determine whether information constitutes personal data, according to the recent interpretation of personal data in case-law \cite{EDPS_SRB}.

%

\noindent \textbf{GDPR obligations for data controllers (app publishers in our context).} In the following, we consider the relevant requirements and principles derived  from both the GDPR and guidelines from the European Data Protection Board (EDPB).\footnote{The EDPB is composed of representatives of all EU Data Protection Authorities (DPAs), EDPB guidelines play a central role in streamlining the interpretation of GDPR.}

Data controllers can be held responsible and fined if they fail to comply with the following obligations under the GDPR~\cite[Art. 28(3)(f), 32-36]{GDPR2018}.
Relevant GDPR principles are presented
below and numbered with P, and legal requirements for consent
revocation are numbered with LR. 

\noindent \textit{- P1 Lawfulness}: when processing personal data, data controllers must collect personal data only with a valid legal basis (Art. 5(1)(a)), such as consent or legitimate interest (LI).
In the context of  online tracking for (non-essential) advertising purposes, consent is the lawful legal basis according to Article 5(3) of the ePrivacy Directive. Further, the GDPR imposes strict requirements for valid consent.
Consent must be i) prior to any data collection, ii) freely given, iii) specific, iv) informed, v) unambiguous, vi) readable, accessible, and vii) revocable (Articles 4(11) and 7 GDPR)~\cite{Sant-etal-20-TechReg}. To be freely given, users must be able to accept or reject consent without facing negative consequences. 

\noindent \textit{- LR1 Correct registration of consent}:
Data controllers must correctly register the user consent revocation decision, and assure that the decision made by the user in the banner interface is identical to the consent that gets registered/stored by the website~\cite[Arts. 7(1), 30, Rec. 42]{GDPR2018},~\cite{Sant-etal-20-TechReg,irish_data_protection_commission_guidance_2025} (p.9). 
A \emph{violation} occurs when a registered consent is different from the user’s choice.

\noindent \textit{- P2 Fairness:} Data controllers must not process personal data in a  unjustifiably detrimental, discriminatory, unexpected or misleading way~\cite[Art.5(1)(a)]{GDPR2018},~\cite[para 17]{edpb_guidelines_2020}.

\noindent \textit{- P3 Transparency}: Data controllers must inform users about purposes, recipients, and legal bases when collecting data (Art. 5(1)(a), 13-14); 

\noindent \textit{- P4 Data Protection by Default}: Data controllers must process data with the highest privacy settings by default (Art. 25(1)(2));

\noindent \textit{- P5 Data Protection by Design}: App developers must implement organizational
measures and technical safeguards efficiently~\cite{EDPB2020-DPbDbDesign}, \cite[Art. 25(1)]{GDPR2018}.


\noindent \textit{- P6 Security}: Data controllers must ensure data protection to prevent unauthorized access (Art. 5(1)(f), 32);


\subsection{Google Advertising ID (AAID) as personal data}
%
\noindent \textbf{Google Advertising ID.} Google Advertising ID, technically identified as \verb|AAID|, is a unique identifier assigned to every Android mobile device.
It is the \textit{de facto} identifier used within the Android ecosystem for  targeted advertising~\cite{google_target_2025}.
Because the AAID is unique to each device, advertisers rely on it to build personalized ad profiles. 
Google’s own Android Developers documentation instructs app developers to restrict the use of this identifier to ``user profiling or ads use cases'', and ``always respect users' selections regarding ad tracking'' when collecting it~\cite{Android-identifiers}.

\noindent \textbf{AAID as personal data.}
The online advertising industry, represented by IAB Europe, explicitly acknowledges that mobile advertising identifiers -- including AAID -- qualify as personal data under the GDPR~\cite{iab_europe_gdpr_implementation_working_group_definition_2018}. This qualification is consistent with GDPR Recital 30, which states that identifiers provided by a user’s device, such as IP addresses or other device identifiers, may be used to single out individuals, thereby making them identifiable. 
The AAID mobile identifier is a direct example of such an ``online identifier''.
It is technically persistent, unique to the device, and is routinely used to track user behavior across apps and services. 
The French Data Protection Authority (CNIL) takes a similar stance, noting that mobile identifiers like the AAID fall under the scope of Article 5(3) of the ePrivacy Directive~\cite{cnil_recommendation_2025}. This means their access and storage require user consent. 





\begin{table}[t]
\centering
\caption[TCF purposes v.2.2, where purposes 1, and 3-6 require consent]{TCF purposes v2.2, where purposes 1, and 3-6 require consent according to the TCF~\cite{tcf-policies}.} 
\label{table:purposes}
\begin{tabular}{|p{0.03\linewidth}|p{0.67\linewidth}|w{c}{.15\linewidth}|}
\hline
Nr & Name & Consent \\
\hline
1 & Store and/or access information on a device & Required \checkmark\\
\hline
2 & Use limited data to select advertising &\\
\hline
3 & Create profiles for personalised advertising & Required \checkmark\\
\hline
4 & Use profiles to select personalised advertising & Required \checkmark\\
\hline
5 & Create profiles to personalise content & Required \checkmark\\
\hline
6 & Use profiles to select personalised content & Required \checkmark \\
\hline
7 & Measure advertising performance &\\
\hline
8 & Measure content performance &\\
\hline
9 & Understand audiences through statistics or combinations of data from different sources &\\
\hline
10 & Develop and improve services &\\
\hline
11 & Use limited data to select content &\\
\hline
\end{tabular} 
\end{table}

\subsection{IAB's Transparency and Consent Framework and Compliance}
\label{subsec:TCF}
The current \textit{de facto} standard for the consent banners in the EU is the IAB Europe Transparency and Consent Framework (TCF). 
The TCF was created in April 2018 with the goal of helping data controllers 
process data according to the legal requirements set by the ePD and the GDPR.
It was developed for use by its participants: i) Publishers of online content where data can be collected, including app publishers, ii) vendors which process user data gathered by the publishers, iii) and Consent Management Platforms (CMPs) which develop consent banners. 
CMPs provide consent management as-a-service, through consent banners embedded in websites and in mobile apps, which request consent and inform users the vendors that can access user data, the purposes these vendors process data for, and the legal bases utilized under these purposes~\cite{TCF}.

\noindent \textbf{Defined purposes.} The current TCF v2.2~\cite{TCF-v.2.2} defines the \emph{pre-defined purposes} for processing personal data by CMPs (presented in  \autoref{table:purposes}), each requiring either consent or legitimate interest as a lawful basis to process data.
%
Versions prior to TCF 2.1 permitted websites and apps to  rely on the legal basis of legitimate interest  for all the purposes for processing data. 
Version 2.1 removed  legitimate interest for Purpose 1 (\textit{Storage or access data to recognise devices}).
The version 2.2 removed legitimate interest from \linebreak personalization-related purposes (Purposes 3-6)~\cite{tcf-policies}.  
Despite these changes, recent work still criticizes  IAB's interpretations of legal grounds for these purposes~\cite{kancherla_johnny_2025}, arguing that  purposes 2 (Use limited data to select advertising), 6 (Use profiles to select personalized content), 7 (Measure advertising performance), 8 (Measure content performance) and 9 (Understand audiences through statistics or
combinations of data from different sources) require consent. 
Accordingly, such purposes should not be enabled by default in the IAB Europe TC string.



\noindent \textbf{Defined format.} The TCF also defines  the \emph{format to store the user's choice}, called TC string. 
When a user interacts with websites or mobile apps which uses the TCF, each website and app will generate a TC string on the user's device containing the user's consent choices. 
This TC string has been qualified as \textit{personal data}~\cite{BelgianDPA2022_Decision21_2022}.
When interacting with websites, the TC strings are stored as HTTP cookies in the user's browser by CMPs. 
On mobile devices, the TC strings are stored locally under two separate strings: \verb|IABTCF_PurposeConsents| and \linebreak \verb|IABTCF_PurposeLegitimateInterests|.
%
Vendors have access to and decode the TC string, and should therefore only process data according to the user's choices~\cite{IAB-API},~\cite{TCF-for-Vendors},~\cite{IAB-TC-String-Storage}.

\noindent \textbf{Version 2.3.} IAB Europe released on June 19th 2025 version 2.3 of the TCF, with a transition period ending on February 28th 2026~\cite{iab_europe_all_2025}.
This new version mostly targets vendors, as it addresses signaling ambiguity. 
It introduces a new field in the TC string (\verb|IABTCF_DisclosedVendors|) to alleviate ambiguity for vendors declaring two types of purposes (``Special Purpose(s) and Purpose(s) under LI'', where a special purpose is a processing purpose for which the user is not given choice by a CMP).
CMPs are also impacted since this new \verb|IABTCF_DisclosedVendors| segment must be disclosed in the user-facing CMP UI.
This new version does not address any of the issues raised in the present article.

\section{Related Work}

This section surveys prior work and shows that earlier TCF versions enabled potential GDPR violations on the Web, that websites frequently exploited the legal basis of legitimate interest  to extract more user data, and that similar compliance shortcomings persist in  Android and iOS mobile apps.

\subsection{Potential GDPR Violations in TCF Research}


Matte et al.~\cite{matte2020cookie, MatteSantosBielova2020} provided one of the earlier accounts of a systematic study of the TCF.
In their empirical work, they crawled 22949 websites, of which 1426 were using the TCF.
They performed extensive tests on 560 of these websites, and concluded that 54.0\% of them were potentially violating both the GDPR and the ePD. 
Violations included websites sharing user consent through “shared cookies”, and storing  consent before giving users a choice, preselected choices, amongst others. 

Smith et al.~\cite{smith2024study} studied in 2023 websites using a more recent version of the TCF (v.2.1).
It is, to the best of our knowledge, the most recent study of GDPR compliance on websites using the TCF. 
The authors denied consent in consent banners of 2~230 websites using TCF-based cookie banners.
Their results show that a majority of the websites were using legitimate interest for purposes 3-6, which are reserved for personalization and subsequently not allowed in TCF v.2.2 (see Section~\ref{sec:background}).
They concluded from a TCF string analysis that 1.3\% of the websites studied had potential GDPR violations, including legitimate interest being used for Purpose 1.
 Morel et al.~\cite{morel2023legitimate} found that that websites using cookie paywalls\footnote{Cookie paywalls allow visitors of a website to access its content only after they make a choice between paying a fee or accept tracking.} always used the TCF. The authors also found websites that deployed legitimate interest to track users 
 even if they chose to pay a fee. 

\subsection{Potential Privacy Violations in Mobile Apps}
\label{sec-previous-violations-mobile-apps}

Nguyen et al.~\cite{nguyen2021share} conducted the first large scale study of potential GDPR violations in Android mobile applications between 2021 and 2022.
The authors initially crawled one million applications over five months, and selected 86000 of these apps for analysis (without banner interaction). 
Using mitmproxy to analyze the network traffic, and Objection to disable SSL pinning (a necessary technical measure preventing apps from expecting communication with a specific host only), they could confirm when applications transmitted their personal data -- such as advertising IDs and location data --, without receiving explicit consent. 
Their results show that 28\% of the studied applications potentially violated the GDPR. 

Koch et al.~\cite{koch2023ok} investigated consent banners between 2022 and 2023 in Android and iOS applications.
This paper furthers the analysis of Altpeter~\cite{altpeter2022ic}, which performed a similar study in 2022.
Altpeter analyzed a dataset of 3421 Android apps, with 2.6\% of apps using the TCF. 
The Android applications studied by Koch et al. were scraped from the top lists of all categories in the Google Play Store, where the authors used the top 100 apps of each category, resulting in a dataset of 3~006 apps. 
They found that 6.6\% of the applications used the TCF. 
After downloading the applications, they performed a static analysis on the preference folder of the app to assess whether a majority of the applications used the same CMPs. 
Other than the TCF-compliant CMPs, most applications used their own solutions to inform users and elicit consent.
They deepened their analysis using Appium (v2.15.0), a software to automate mobile app testing.\footnote{\url{https://appium.io/docs/en/latest/}}
They were able to classify how the apps presented their consent banner when launched, where the banners were classified as 1) links, 2) notices, or 3) proper banners. 
They observed that of the three alternatives, only the proper banners required users to interact by consenting or denying data collection before using the app, while the others stated that data will be collected if the user continues to use the app. 
Lastly, the authors analyzed the apps' traffic, similarly to the methods presented by Nguyen et al.~\cite{nguyen2021share}. 
The analysis was performed by comparing the data transmitted by applications with what the authors had consented to in the applications. 
The results show that 43.1\% of total studied apps on Android and iOS potentially violated the GDPR.
No work had previously systematically studied the privacy implications of the TCF on Android applications.

\begin{figure*}[!ht]
    \centering
    \caption{Schematic overview of the three steps of the methodology.
     First, we quantify the TCF in a dataset of 4482 apps scraped on the Play Store (Section~\ref{subsec:quantification}). Second, we automate the interaction with consent banners on 576 apps according to three approaches (Section~\ref{subsec:dynamic_analysis}). Third, we analyze network traffic of up to 550 apps in two stages (Section~\ref{subsec:traffic-analysis}).} 
    \includegraphics[scale=.9]{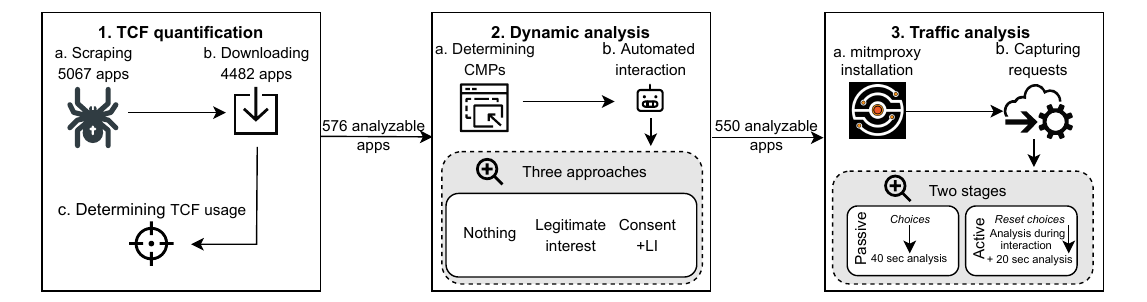}

    \label{fig:methodology}
\end{figure*}

\section{Methodology}
We describe in this section our methodology.
We begin by explaining in Section~\ref{subsec:quantification} how we scraped the Google Play Store and downloaded apps to quantify the prevalence of the TCF, addressing \textbf{RQ1}. 
We then describe in Sections~\ref{subsec:dynamic_analysis} and~\ref{subsec:traffic-analysis} how we analyzed TCF-based applications through dynamic and traffic analysis, answering \textbf{RQ2} and \textbf{RQ3} respectively. 
A visual overview is provided in Figure~\ref{fig:methodology}.

\noindent \textbf{Scope and technical setup.}
This paper only targets TCF-based Android applications, and no applications on other operating systems such as iOS. 
We crawled the top apps of each category (see also Appendix~\ref{sec:ethics} for the rationale behind this technical choice) of the Google Play Store (also called Google Play or Play Store) until we collected at least 5000 applications (66\% more than the state of the art~\cite{koch2023ok} for a similar type of analysis) between 2025-01-29 and 2025-03-10.
All results are based on data gathered from applications in this dataset.
We believe our sampling set is representative of Android applications because it represents popular and commonly used apps.
We accessed the Google Play Store with a fixed EU vantage point (verified between app launches), ensuring that EU data protection laws (\textit{i.e.}, the GDPR and the ePD) apply.
We used a newly created Gmail account dedicated to the project and a phone number automatically generated by the emulator.
To avoid inter-app interaction (during the dynamic and traffic analysis), apps were deleted between testing two of them once their analysis was completed.


\subsection{Quantifying the Presence of the TCF}
\label{subsec:quantification}
\subsubsection{Scraping Apps} 
We first created a dataset consisting of application package names, application names, application category and the date of scraping. 
We used Selenium\footnote{\url{https://www.selenium.dev/}} (v4.28.1), a software used to automate browser interactions.
Our Selenium instance was not logged in our Google account while crawling, and used the same IP and location throughout the scraping.
The Play Store possesses top charts for each category: \nolinkurl{https://play.google.com/store/apps/category/CATEGORY} which contain between 60 and 900 applications depending on the category. 
The Play Store offers 49 categories\footnote{\url{https://www.searchapi.io/docs/parameters/google-play-store/categories}}, or 56 with the \textit{family} apps (targeting children).
We scraped all categories daily between 2025-01-29 and 2025-03-10, until we obtained 5067 unique applications.

\subsubsection{Downloading Apps} 
We then instantiated an emulated Android device through Android Studio\footnote{\url{https://developer.android.com/studio}}, allocated it with 300 GB of storage and gave the device superuser privileges by using Magisk\footnote{\url{https://magiskmanager.com/\#What_is_Magisk}}, a software used to root Android devices. 
The device was selected to be a Google Pixel 9 Pro using API 35. 
Google Pixel 9 Pro was the newest available device in Android Studio at the time of the study, and API 35 was a recent stable release. 
We created an ad-hoc solution to download applications, building upon the Android Debug Bridge\footnote{\url{https://developer.android.com/tools/adb}} (\verb|adb|), a Unix shell allowing for interaction with emulated and real Android devices. 
Since the Play Store application worked on the emulated device, we were able to automatically download applications. 
When launching the Play Store from the emulated device through \verb|adb|, it was possible to provide an app package name to be directly redirected to the app's page in the Play Store and proceed with downloading the application.



\subsubsection{Determining Apps' TCF Usage} 
According to the IAB's CMP API, applications using the TCF should store their TC Strings (such as \verb|IABTCF_TCString|) in files stored in the \verb|SharedPreferences| directory\footnote{SharedPreferences is a directory which typically contains files composed of collections of key-value pairs for an Android application.}~\cite{IAB-API}. 
We launched and closed apps using the \verb|adb| commands \verb|monkey|~\footnote{A command-line tool ran on any emulator instance or on a device, it sends a pseudo-random stream of user events into the system.} and \verb|am force-stop|~\footnote{To force-stop an activity. 
In an Android context, an activity is a window where an app has its UI, activities may sometimes float over another activity, such as pop-ups cf \url{https://developer.android.com/guide/components/activities/intro-activities}} to generate the \linebreak \verb|SharedPreferences| directory of applications. 
The script then executed the Unix-command \verb|grep| (a utility for searching patterns in a file) on all files in the apps' generated  \verb|SharedPreferences| directory, searching for mentions of \verb|IABTCF|.   

However, apps could well initialize the framework without implementing a banner, we therefore confirmed the presence of the TCF by restricting our data analysis to apps able to present a banner.
The previous steps nonetheless provided an initial pruning of the data to a restricted set of candidate apps.


\begin{figure*}
    \centering
    \begin{subfigure}[b]{0.4\textwidth}
        \includegraphics[width=\textwidth, keepaspectratio, trim = {31.2cm, 6cm, 25.7cm, 5.7cm}, clip]{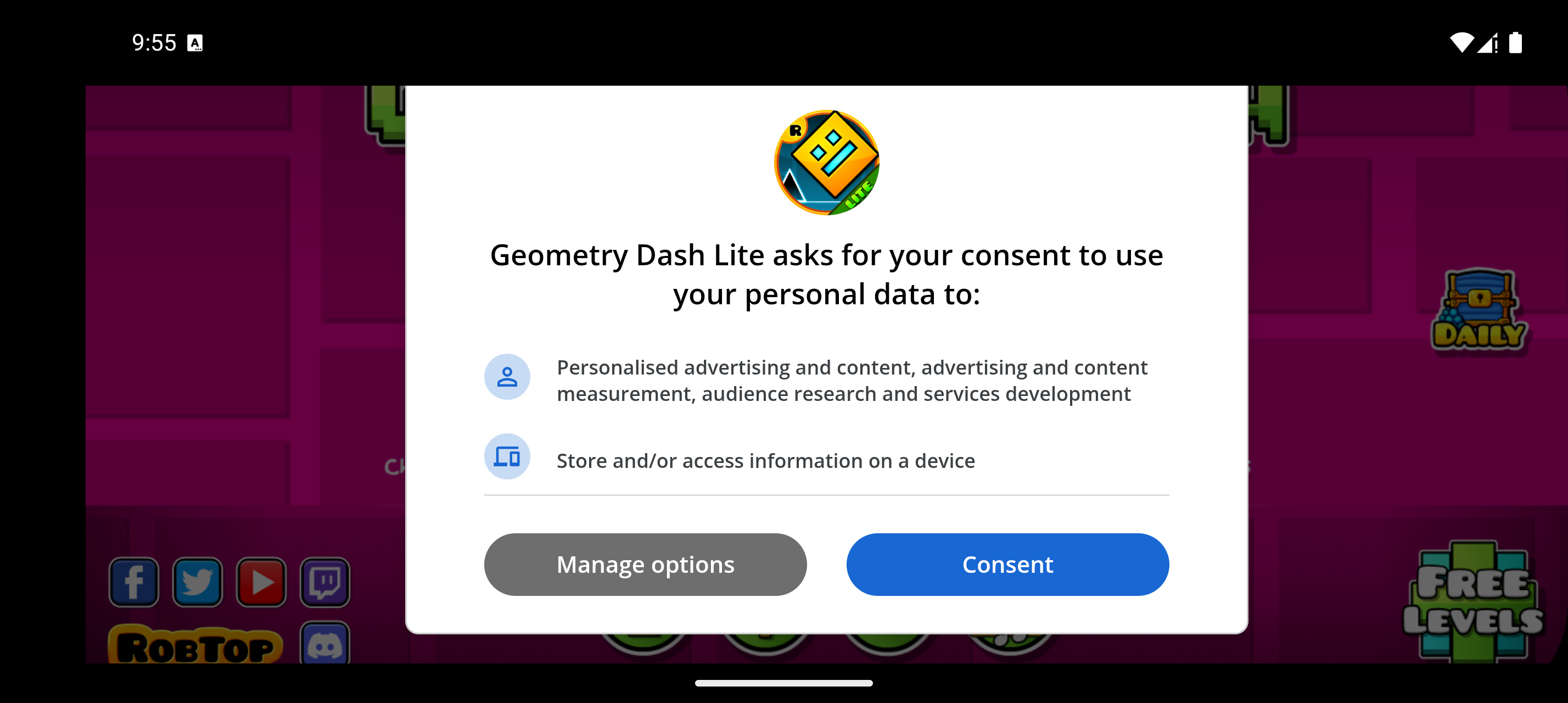}
        \caption{First layer of the consent banner.}
        \label{fig:300_first}
    \end{subfigure}
    \qquad
    \begin{subfigure}[b]{0.4\textwidth}
        \includegraphics[width=\textwidth, keepaspectratio, trim = {31.2cm, 6cm, 25.7cm, 5.7cm}, clip]{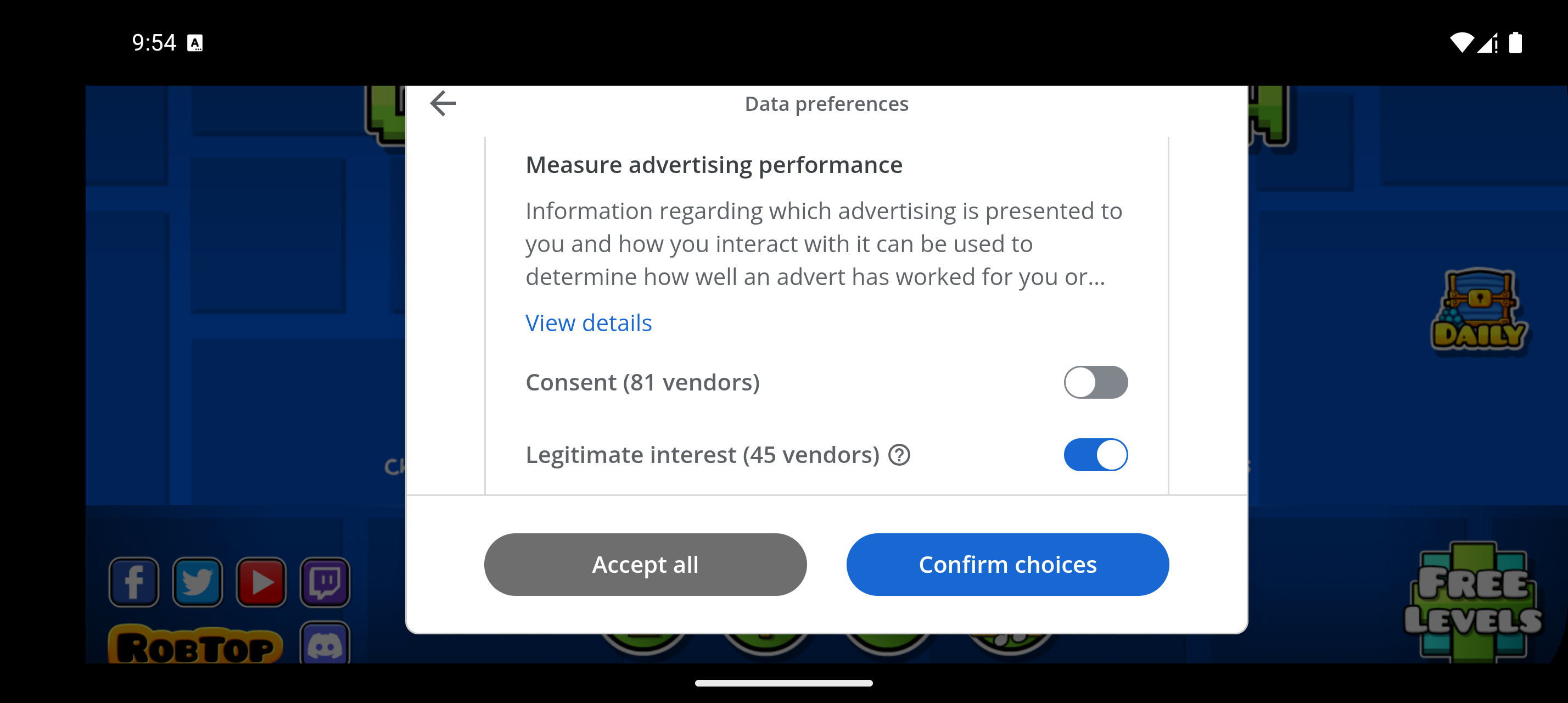}
        \caption{Detail of a purpose. All LI purposes are pre-checked.}
        \label{fig:300_purpose}
    \end{subfigure}
    \caption[Consent banner settings for Google LLC (CMP 300), from the application Geometry Dash Lite]{Consent banner settings for Google LLC (CMP 300), from the application Geometry Dash Lite (com.robtopx.geometryjumplite). The two screenshots show default interfaces.}
    \label{fig:cmp-300-options}
\end{figure*}

\subsection{Dynamic Analysis}
\label{subsec:dynamic_analysis}
To assess whether TCF-based apps correctly store users' choices, we performed a dynamic analysis using \verb|adb| and Appium, as the latter also enables an automatic launch of apps and interaction with their consent banners. 
Since the design of consent banners are based on the CMP used, we therefore initially researched the most common CMPs in our dataset to streamline our analysis. 

\subsubsection{Determining CMPs}
In a majority of applications, the CMP could simply be found in the app's \verb|preferences.xml| file in its \linebreak \verb|SharedPreferences| directory by looking at the string \linebreak \verb|IABTCF_CmpSdkID|. 
If not, we manually launched the application and inspected its consent banner for patterns. 
Some applications had their CMP mentioned in other files in their \linebreak \verb|SharedPreferences| directory, and some applications had their CMP included in their application package name, such as \linebreak \verb|com.outfit7.mytalkingtom2| where the CMP is Outfit7.
After having obtained the CMP of each application, we manually inspected some consent banners of each CMP to establish cases later used to automate the interaction via Appium.
A majority of cases consisted of waiting for certain elements to show up on the mobile screen, and then scrolling through the consent banner to interact with buttons and toggle-buttons based on the buttons' content-description- or text-elements. 

\subsubsection{Automated interaction}
\label{subsec:auto_interaction}
To determine which elements we interacted with, we first manually investigated consent banners from each CMPs to find regularly occurring sentences/words on target buttons, we then used those recurring sentences as \textit{target sentences}. 
These target sentences were used as input into a Python function using the multilingual sentence similarity model \linebreak \verb|paraphrase-multilingual-MiniLM-L12-v2|~\cite{reimers-2019-sentence-bert,sentence-transformer-model}.
The function also takes a list of expected texts and then outputs the element most likely to be the one we want to interact with according to a threshold. 
The model created was later passed as instructions to Appium. 
If our automated interactions failed, \textit{i.e.}, we failed to interact with buttons in a banner, we would manually look at the banner in question and potentially add the sentences from the app’s buttons as target sentences, thereby refining the model.
We validated our automated banner interactions by confirming on a random sample that the \verb|SharedPreferences| of the sampled apps had the expected values (\textit{e.g.}, no purposes in use if our automatic interaction was denying all purposes).


The most commonly used CMP in our dataset is Google LLC (CMP 300, see Table~\ref{tab:CMPs}), a screenshot is presented as an example in~\autoref{fig:cmp-300-options}. 
This CMP allowed us to interact with consent and legitimate interest in a standardized manner after clicking ``Manage option''. 
Figure~\ref{fig:300_purpose} shows a detail of the purpose ``Measure advertising performance'' and how it includes both legitimate interest and consent, with consent following an opt-in basis and legitimate interest an opt-out basis. 
Other CMPs provided less standardized interfaces (the UI could greatly vary across different implementations), or had various extra steps to reach the options containing consent and legitimate interest, requiring additional manual tuning.

\subsubsection{Banner Interaction Approaches} 
\label{subsec:approaches}
We interacted with consent banners through three different approaches: 
\begin{enumerate}
    \item  agree to all, denoted \textbf{C+LI},
    \item  legitimate interest purposes only, denoted \textbf{LI},
    \item  disagree to all, denoted \textbf{Ø}. 
\end{enumerate}
The first approach enables a broad analysis of the apps' traffic, determining what personal data is being transmitted to different servers when permission is given by the user. 
The rationale behind the \textbf{LI}-approach was to analyze whether any legitimate interest purposes forbidden by the TCF were actually in use, while the \textbf{Ø}-approach should yield a lower bound for data collection.
We also used the \textbf{Ø}-approach to analyze the traffic of applications, and to compare the data transmitted without consent versus the data transmitted with consent to all (\textbf{C+LI}-approach). 
An approach where we provide consent without legitimate interest would be meaningless, as this has no functional difference from the \textbf{C+LI}-approach where we agree to everything. A visual representation of the three different approaches can be seen in~\autoref{table:interaction-approach-explanation}.

\begin{table}[t]
\centering
\caption{Purposes accepted according to their legal bases during each interaction approach.} 
\label{table:interaction-approach-explanation}
\begin{tabular}{|c|c|c|}
\hline
\textbf{Approach} & \textbf{Consent} & \textbf{Legitimate Interest}\\
\hline
\textbf{Agree to all (C+LI)} & \checkmark & \checkmark\\
\hline
\textbf{Leg. Int. purposes (LI)} & \textbf{X} & \checkmark\\
\hline
\textbf{Disagree to all (Ø)} & \textbf{X} & \textbf{X}\\
\hline
\end{tabular} 
\end{table}

All three approaches required interacting with each CMP in three different ways. 
The \textbf{C+LI}-approach often solely required pressing an ``accept all''-button. 
The \textbf{LI}-approach demanded locating the button allowing a manual change in our banner choices, followed by a confirmation of the choices automatically provided (since legitimate interest is on an opt-out basis).
The \textbf{Ø}-approach required manually changing our choices, and then opting-out of every purpose related to legitimate interest. 

\subsubsection{Confirming Stored Banner Choices} 
Upon interaction with consent banners, we confirmed the choices stored by the applications in their \verb|preferences.xml| file.
We leveraged the Unix-command \verb|grep|, parsing relevant strings such as \linebreak \verb|IABTCF_PurposeConsents| and \linebreak \verb|IABTCF_PurposeLegitimateInterests|.
These strings act as two TC strings (one for each legal ground), and should therefore reflect users choices.
We stored the results from our interactions with consent banners in CSV files.

\subsection{Traffic Analysis}
\label{subsec:traffic-analysis}
We captured network traffic of apps in our dataset to evaluate their respect of user choices (disagree to all, legitimate interest only, agree to all), and analyzed eventual personal data collection.
We leveraged the mitmproxy suite\footnote{Suite of tools that provides an interactive HTTPS-capable interceptor proxy: \url{https://mitmproxy.org/}} (v11.1.3) to capture traffic, and Objection (v.1.11.0) to disable SSL pinning\footnote{Method used by Android apps to prevent man-in-the-middle attacks: \url{https://www.indusface.com/learning/what-is-ssl-pinning-a-quick-walk-through/}}. 
Objection is a toolkit powered by Frida~\cite{frida-homepage}, which therefore additionally required installing a Frida server on our device. 


\subsubsection{Traffic Analysis Preparation}
We initialized our setup by installing and trusting mitmproxy's certificate authority~\cite{mitmproxy-certificate-authority}, allowing mitmproxy to decrypt encrypted traffic.
The certificate authority was downloaded on our device through a Magisk module~\cite{magisk-module}. 
We ensured that our device had a European IP and that the device's GPS coordinates were set in an EU country to ensure that the GDPR and the ePD applied to our device. 
These coordinates could then be seen in the device's fused location, which is described by Google as a combination of underlying location technologies, such as GPS and Wi-Fi~\cite{fused-location}. 
We accessed the device's coordinates by using the \verb|adb|-command \verb|dumpsys location|. 

\subsubsection{Traffic Analysis Workflow}
\label{sec:workflow}
We captured traffic by combining the mitmproxy functionality \verb|mitmdump| with a Python script of our choice as a mitmproxy addon, allowing for customization of mitmproxy's behavior. 
The add-on running alongside \linebreak \verb|mitmdump| searched through the payload and URL of each HTTP-request on our emulated device for personal data. 
To identify personal data~\footnote{\label{ft:pd} We focused on unique identifiers, since the definition of personal data in the GDPR leaves little room for interpretation: ``an identifier such […] an identification number, location data, \textbf{an online identifier}'' (emphasis is ours).} in requests, we used a) hard-coded strings, b) strings based on running \verb|adb| commands to extract present data on the device -- such as \verb|AAID| and location data --, and c) regular expressions for data such as the device's phone number and MAC address, due to the different formats they could be presented in. 
Informed by the literature, we use the following list: \textit{AAID, public IP, email, phone number, IMEI, MAC address, GPS location, IMSI, ICCID and phone serial number}.
If any personal data was found in a request, we stored the following in a Postgres-database: the application package name, the personal data being transmitted and the request method, the request-url, the type of personal data being transmitted, and the date of the request.


%
We then launched each application through Objection with the startup-command \verb|android sslpinning disable| -- which allowed for traffic collection in apps --, and captured traffic before launching the next one (according to two stages further described in Section~\ref{subsec:stages}).
Certain applications would detect SSL pinning being disabled and crash.
We therefore confirmed the running of the application by monitoring processes on the device. 
If the process related to the current application was preemptively terminated, we re-executed the app up to two times, before labeling the app as non-analyzable and moving on to the next application.  

\subsubsection{Device Snapshots} 
\label{subsec:snapshots}
The crash of an app would occasionally cause our device to crash or reboot, slowing down the analysis.
We took regular snapshots of our device to work around this mishap, preserving the device's state at the time of the snapshot~\cite{android-snapshots}.
Using snapshots was much faster than rebooting the phone and was a necessity as the crashes were frequent during our traffic analysis. 
We created a snapshot for each dynamic analysis approach, capturing the application state at the point when our banner choices had been stored for that specific approach.

\subsubsection{Stages of Traffic Analysis} 
\label{subsec:stages}
In addition to capturing traffic during the three approaches previously mentioned, we analyzed apps' traffic in two stages, called \textit{passive}- and \textit{active}-stages. 

\textit{Passive}-stage: we analyzed apps' traffic for 40 seconds, while the apps remained passive, when the banner choices were already stored. 
The rationale of this stage was simply to observe the traffic \textit{ex-post} to  the user choices communicated in the banner, without any additional interaction with the banner from our side (hence, the name \textit{passive}).

\textit{Active}-stage: we reset all banner choices, and then analyzed traffic during display or navigation of the apps' consent banners (before the user confirms consent choices), and also while idling for 20 seconds after a choice was made. 
The rationale behind this second stage was to enable observation of the network traffic \textit{during} the interaction with consent banners (hence, the name \textit{active}).

\section{Results}
\label{sec:results}
This section  presents the number of applications scraped and the percentage of them using the TCF, answering \textbf{RQ1}. 
It then introduces the results of our dynamic analysis, assessing the registration of banner choices, addressing \textbf{RQ2}. 
Last, it presents the results of our traffic analysis, across the three approaches (described in Section~\ref{subsec:approaches}) and the two stages (described in Section~\ref{subsec:stages}), answering \textbf{RQ3}. 
The raw data used to produce the results can be found in our public anonymized repository~\cite{public-repo}.



\subsection{Quantifying the TCF}
\label{sec:result-rq1}
The Play Store was scraped daily, between 2025-01-29 and 2025-03-10, resulting in an initial dataset of 5067 apps.
Downloading was performed between 2025-02-04 and 2025-03-12.
585 of the 5067 apps presented issues during downloading, for a total of 4482 downloaded apps.
Some apps were premium and required a payment to install, a few apps were pre-registrations and did not provide an installable option at the time, and some applications were removed from the Play Store between the creation of our dataset and our attempted download.
The most prevalent reason was the detection of an emulated device, the installation button was then replaced with the message: ``Your device isn't compatible with this version.''.
We did not record the distribution of the failure analysis.


\subsubsection{Google is by far the main CMP  found in TCF-based apps.}
We encountered twelve different CMPs in our dataset, presented in Table~\ref{tab:CMPs}. 
The most common CMP in the dataset was Google LLC with a large majority, amounting to \textbf{89.76\%} of apps.
The distribution follows a power law distribution, with one CMP being clearly more prevalent than the others, and a long tail of five CMPs amounting to less than 10\% of the total, indicating an extremely unbalanced distribution amongst CMPs.

\begin{table}[]
\caption{Consent Management Platforms (CMPs) found in TCF-based applications, including their prevalence. The ID is used by the TCF to identify CMPs.}



\label{tab:CMPs}

\begin{tabular}{|c|l|c|c|}
\hline
\textbf{ID} & \textbf{Name} & \textbf{Number of apps} & \textbf{Proportion} \\ \hline
300 & \textbf{Google LLC} & \textbf{517} & \textbf{89.76\%} \\ \hline
5 & Usercentrics & 16 & 2.78\% \\ \hline
350 & Easybrain Ltd & 14 & 2.43\% \\ \hline
No id & Data Unavailable & 11 & 1.90\% \\ \hline
348 & Outfit7 Limited & 9 & 1.56\% \\ \hline
7 & Didomi & 7 & 1.22\% \\ \hline
28 & OneTrust LLC & 2 & 0.35\% \\ \hline
\end{tabular}

\end{table}

\begin{table}[H]
\caption{Top 10 TCF-based app  categories}
\label{tab:top_10_categories}

\centering
\begin{tabular}{|p{0.45\linewidth}|p{0.45\linewidth}|}
\hline
\textbf{App Category} & \textbf{TCF usage}\\
\hline
PERSONALIZATION & 54/101 (53.47\%) \\
\hline
LIBRARIES\_AND\_DEMO & 26/76 (34.21\%) \\
\hline
ART\_AND\_DESIGN & 27/79 (34.18\%) \\
\hline
VIDEO\_PLAYERS & 20/67 (29.85\%) \\
\hline
TOOLS & 26/90 (28.89\%) \\
\hline
PHOTOGRAPHY & 22/77 (28.57\%) \\
\hline
MAPS\_AND\_NAVIGATION & 22/86 (25.58\%) \\
\hline
PRODUCTIVITY & 16/67 (23.88\%) \\
\hline
MUSIC\_AND\_AUDIO & 20/85 (23.53\%) \\
\hline
WEATHER & 15/65 (23.08\%) \\
\hline
\end{tabular}
\end{table}

\subsubsection{TCF-based apps can be found in all Play Store categories and many countries outside the EU} 
\label{subsec:cat_count}
As we scraped the Play Store per app category, we were able to provide the first quantification of TCF-based Android apps per category and country. 
The most common app categories using the TCF in our dataset are: Personalization (53.47\%), Libraries and Demo (34.21\%), and Art and Design (34.18\%) (see Table~\ref{tab:top_10_categories}, and Table~\ref{appendix:categories} for the comprehensive data).

We additionally quantified the country of origin of apps in our dataset.
We found that \textbf{82.14\% of apps were developed in non-EEA countries}. 
Only one country (Cyprus) in this top 10 is part of the EU/EEA (see Table~\ref{tab:top_10_countries}), with three other adequate in the sense of GDPR Article 45 according to the EU Commission~\cite{eu_commission_data_2025} (USA, Israel, UK).
See Table~\ref{appendix:developer-country-statistics} in the Appendix for the comprehensive data.

\paragraph{\textbf{Summary:}}
In total, 576 apps were identified as implementing the TCF (842 initialize the framework, but 576 were confirmed to use the TCF), representing 12.85\% of the downloaded apps.
TCF-based apps are widely spread across categories, although Personalization has a significant lead (53.47\%, vs. 34.21\% for the runner-up).
Most of the TCF-based apps are developed outside the EU/EEA or adequate countries (in the sense of GDPR Article 45).

\begin{table}[H]
\caption{Distribution of the top 10 countries where TCF-based apps in our dataset were developed}

\label{tab:top_10_countries}
\centering
\begin{tabular}{|p{0.45\linewidth}|p{0.45\linewidth}|}
\hline
\textbf{Country} & \textbf{Number of TCF-based apps in our dataset (N = 576)}\\
\hline
Vietnam & 94 (16.32\%) \\
\hline
Hong Kong & 60 (10.42\%) \\
\hline
Cyprus & 41 (7.12\%) \\
\hline
China & 41 (7.12\%) \\
\hline
Israel & 36 (6.25\%) \\
\hline
India & 36 (6.25\%) \\
\hline
United States & 27 (4.69\%) \\
\hline
Pakistan & 22 (3.82\%) \\
\hline
United Kingdom & 17 (2.95\%) \\
\hline
United Arab Emirates & 15 (2.6\%) \\
\hline
\end{tabular}
\end{table}

\begin{table*}[!ht]
\centering
\caption{Traffic analysis results per phase of interaction. The last column contains the \textbf{active}-stage results amended with data from apps sharing AAID during consent banner interaction OR during a given approach.}
\label{table:traffic-amended-results}
\begin{tabular}{|c|c|c|c|}
\hline
\textbf{Banner Interaction} & \textbf{Passive-stage (N~=~550)} & \textbf{Active-stage (N~=~513)} & \textbf{Active-stage Amended (N~=~513)} \\
\hline
\textbf{LI}-Approach & 351 (63.8\%) & 192 (37.4\%) & 339 (66.1\%)\\
\hline
\textbf{Ø}-Approach & 364 (66.2\%) & 145 (28.2\%) & 330 (64.3\%)\\
\hline
\textbf{C+LI}-Approach & 367 (66.7\%) & 217 (42.3\%) & 347 (67.6\%)\\
\hline
During Interaction & - & 284 (55.3\%) & -\\
\hline
\end{tabular} 
\end{table*}



\subsection{Dynamic Analysis}
\label{sec:result-rq2}

Regardless of the interaction method, 
none of the 576 analyzed apps used legitimate interest-based purposes forbidden by the TCF (see Section~\ref{subsec:TCF}).
However, 15 apps only stored our choices if provided with consent to all data processing, the user would otherwise be prompted with a consent banner to provide consent again on the next launch of the app:
\textit{Sweet Beauty Camera Filter, Geotag Photo: Camera Location, Control Center Simple, Draw Animation - Anim Creator, Silly Smiles Live Wallpaper 4K, AR Draw Sketch: Trace \& Paint, Easy Piano Keyboard - Piano88,  Find Phone by Clap, Whistle, Caller ID Name, Location \& SMS, Chat AI - AI Chatbot Friend, Dating - Chat \& Meet Singles, Locate Mobile by Number, Phone Tracker By Number, Books - Read and Download, PDF Viewer: PDF Fill \& Sign} (see Table~\ref{tab:incorrec-registration} for the package names). 

\paragraph{\textbf{Summary:}}
In total, \textbf{15 apps only stored our choices if provided with consent to all data processing}, but our analysis did not capture apps misimplementing TCF purposes.




\subsection{Traffic Analysis}
\label{sec:result-rq3}

We detail the prevalence of personal data (see Footnote~\ref{ft:pd}) shared by TCF-based apps in our dataset to third-party domains (see Section~\ref{sec:domains}), according to the two stages of analysis performed (\textit{passive} and \textit{active}) and across the three interaction approaches (\textbf{Ø}, \textbf{LI}, and \textbf{C+LI}).
The only types of data captured in our analysis are \verb|AAID| and IP addresses. 
Since an IP address is also collected for communication purposes in order to initiate HTTP requests, we only present results regarding AAID collection.
A summary of our results can be found in~\autoref{table:traffic-amended-results}.
Assuming that our sample is representative of TCF-based apps, \textbf{up to 66\% of apps share AAID without consent}. 


\subsubsection{Passive-stage} 
Of the 561 apps that we could interact with in the dynamic analysis, 550 apps (98.03\%) had their traffic analyzed during the \textit{passive}-stage between 2025-03-23 and 2025-05-17. 
11 apps could not have their traffic analyzed as they would instantly crash or deny SSL pinning being disabled. 
%
Using the \textbf{LI}-approach, 351 applications were found to transmit \verb|AAID|. 
Using the \textbf{Ø}-approach, 364 applications were found to transmit \verb|AAID|. 
Using the \textbf{C+LI}-approach, 367 applications were found to transmit \verb|AAID|. 





\subsubsection{Active-stage} 
We obtained slightly different results for the \textit{active}-stage, during which we analyzed the traffic before and while interacting with apps' consent banners. 
%
This analysis was conducted between 2025-05-15 and 2025-05-21, during which we inspected network traffic for 513 apps (91.44\% of the apps analyzed in Section~\ref{subsec:dynamic_analysis}). 
37 additional apps could not have their traffic analyzed in this stage since they had been removed from the Play Store, no longer presented a consent banner, or instantly crashed on launch. 
During this stage, we found 284 apps that transmitted  \verb|AAID| before or during banner interaction, prior to obtaining consent. 
%
%
Using the \textbf{LI}-approach, 192 applications were found to transmit \verb|AAID|. 
Using the \textbf{Ø}-approach, 145 applications were found to transmit \verb|AAID|. 
Using the \textbf{C+LI}-approach, 217 applications were found to transmit \verb|AAID|.


\subsubsection{Domains Receiving AAID} 
\label{sec:domains}
\begin{figure}
    \centering
    \includegraphics[scale=.28]{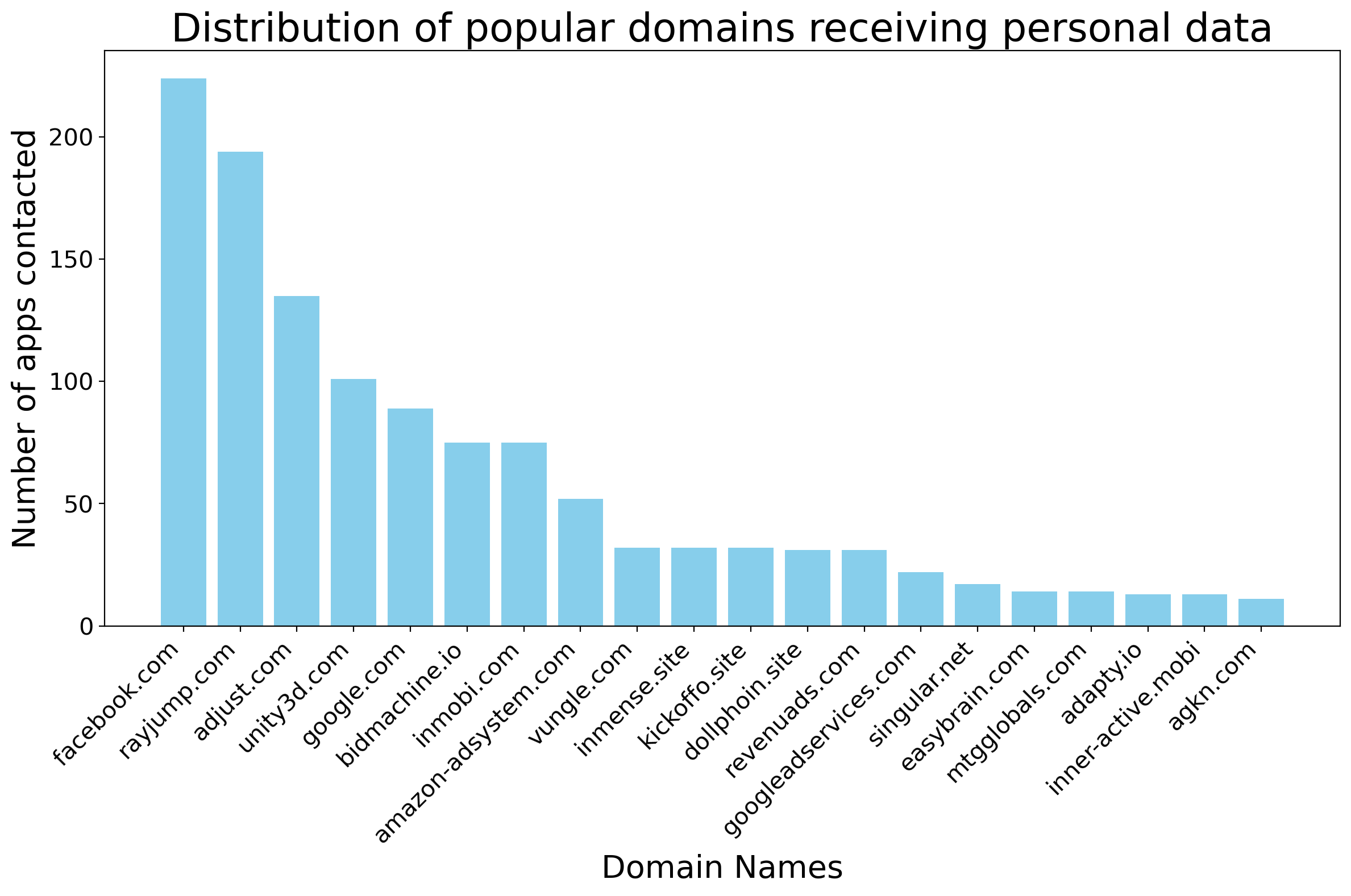}
    \caption{Distribution of popular domains (minimum 10 apps) receiving personal data.} 
    \label{fig:domains}
\end{figure}

We collected the domains (as a combination of SLD and TLD, we do not differentiate subdomains in our analysis) most often contacted with AAID across all interactions approaches per app (see Figure~\ref{fig:domains}).
The five most common domains receiving AAID are: 
\begin{enumerate}
    \item Facebook.com
    \item Rayjump.com\footnote{Note that rayjump.com is widely seen as a scam/malicious website~\cite{meskauskas_rayjumpcom_2024, anyrun_malware_2025}.}
    \item Adjust.com
    \item Unity3D.com
    \item Google.com
\end{enumerate}

\subsubsection{Games are the top app categories sharing AAID without consent} 
\label{subsec:games}
A cross-examination on the categories of apps (combined passive and active stages, on both \textbf{LI}- and \textbf{Ø}-approaches) yields that game apps are the top apps sharing AAID without consent.
Table~\ref{tab:top_categories_traffic} presents the top 10 categories of apps that share AAID without consent, see Table~\ref{tab:cross_categories_sharing} for the comprehensive data.

\begin{table}[h]
\caption{Top 10 categories of apps that shared AAID (combined passive and active stages, on both \textbf{LI}- and \textbf{Ø}-approaches)}
\label{tab:top_categories_traffic}

\centering
\begin{tabular}{|p{0.68\linewidth}|p{0.22\linewidth}|}
\hline
\textbf{Category} & \textbf{Violation \%} \\
\hline
GAME\_SIMULATION & 12/12 (100.0\%) \\
\hline
GAME\_ARCADE & 12/12 (100.0\%) \\
\hline
GAME\_SPORTS & 11/11 (100.0\%) \\
\hline
GAME\_CASUAL & 10/10 (100.0\%) \\
\hline
GAME\_CASINO & 4/4 (100.0\%) \\
\hline
EDUCATION & 3/3 (100.0\%) \\
\hline
AUTO\_AND\_VEHICLES & 2/2 (100.0\%) \\
\hline
GAME\_BOARD & 12/13 (92.31\%) \\
\hline
GAME & 23/26 (88.46\%) \\
\hline
PRODUCTIVITY & 14/16 (87.5\%) \\
\hline
\end{tabular} 
\vspace{5mm}
\end{table}

\subsubsection{Results per developer and per CMP}
\label{subsec:per_dev}

We grouped our results per developer to investigate whether developers adopt the same practices across their different apps.
The developer name and contact is provided on an app's webpage in the Play store.

The vast majority of developers either \textit{always} provide apps sharing AAID without consent, \textit{i.e.}, all their apps share AAID in either the \textbf{LI}- and \textbf{Ø}-approach (238 developers in the passive stage, 221 in the active stage), or \textit{never}, \textit{i.e.}, none of their app share AAID (131 in the passive stage, 146 in the active stage).
Only a handful of developers have mixed results (12 in the passive stage and 14 in the active stage), see Tables~\ref{tab:violating_developers_stage_1} and~\ref{tab:violating_developers_stage_2}.
This result confirms that apps behave differently across developers, although the number of ``rogue'' developers is high, excluding the hypothesis of a few developers making up the majority of ``bad'' apps.

We also noted that \textit{the two top rogue developers in the passive stage also happen to be CMPs} (EasyBrain Ltd and Outfit7 Limited, with EasyBrain Ltd also having a 100\% rate in the active stage).
Further examination into the proportion of apps sharing AAID without consent per CMP yields that these two CMPs were often used by apps sharing AAID without consent (see Tables~\ref{tab:violating_CMP_stage_1} and~\ref{tab:violating_CMP_stage_2}).





\begin{table}[t]
\caption{Violating apps per CMP in passive stage}
\label{tab:violating_CMP_stage_1}

\centering
\begin{tabular}{|p{0.60\linewidth}|p{0.30\linewidth}|}
\hline
\textbf{CMP} & \textbf{Violation \%} \\
\hline
Easybrain Ltd & 14/14 (100.0\%) \\
\hline
Outfit7 Limited & 9/9 (100.0\%) \\
\hline
Usercentrics & 15/16 (93.75\%) \\
\hline
Google LLC & 323/517 (62.48\%) \\
\hline
Didomi & 4/7 (57.14\%) \\
\hline
Onetrust LLC & 1/2 (50.0\%) \\
\hline
Data Unavailable & 0/11 (0\%) \\
\hline
\end{tabular}
\end{table}

\begin{table}[t]
\caption{Violating apps per CMP in active stage}
\label{tab:violating_CMP_stage_2}

\centering
\begin{tabular}{|p{0.60\linewidth}|p{0.30\linewidth}|}
\hline
\textbf{CMP} & \textbf{Violation \%} \\
\hline
Easybrain Ltd & 14/14 (100.0\%) \\
\hline
Outfit7 Limited & 7/9 (77.78\%) \\
\hline
Usercentrics & 12/16 (75.0\%) \\
\hline
Google LLC & 303/517 (58.61\%) \\
\hline
OneTrust LLC & 1/2 (50.0\%) \\
\hline
Didomi & 3/7 (42.86\%) \\
\hline
Data Unavailable & 0/11 (0\%) \\
\hline

\end{tabular}
\end{table}



\paragraph{\textbf{Summary:}}
In the \textit{passive}-stage of traffic analysis, \textbf{66.2\%} of analyzed apps share  personal data when using the \textbf{Ø}-approach. 
During the \textit{active}-stage of analysis, \textbf{55.3\%} of apps also share personal data before the user confirms consent choices (\textit{i.e.}, during banner display or navigation).
Based on these observations, \textbf{the majority of the surveyed TCF-based Android apps do not respect users' banner choices at runtime and collect personal data regardless of users choices}.
We also observe that games are the categories most impacted by potential violations, and that developers often tend to have homogenous practices amongst their apps (\textit{i.e.}, either always or never sharing AAID without consent).


\section{Discussion}

In this section, we contextualize our findings within prior work and discuss the implications of our results.
First, in Section~\ref{subsec:context}, we situate our results in relation to existing research. We then connect our findings to our research questions in Sections~\ref{subsec-discussion-results-rq1} to~\ref{subsec:discussion_games}.
Alongside, we examine the legal implications of our results 
in light of the data protection requirements outlined in Section~\ref{subsec:gdpr-epd}. 
We outline the limitations of our research in Section~\ref{subsec:limitations} and identify research avenues for future work in Section~\ref{subsec:future}.

\subsection{Contextualization and implications}
\label{subsec:context}

\subsubsection{Growing prevalence of the TCF}
\label{subsec-discussion-results-rq1}
Our results show that 12.85\% of downloaded apps  use the TCF. 
The most recent work on TCF-based Android apps -- Koch et al.~\cite{koch2023ok} -- reported in 2023 that 6.6\% of the apps in their dataset implemented the TCF. 
This already reflected a substantial rise compared to Altpeter's~\cite{altpeter2022ic} findings, who found only 1.9\% of apps in 2022. 
%
Our findings of 12.85\% indicates that  TCF adoption  \textit{in Android apps has grown by a factor of 1.95 over the past two years} compared to previous work, although part of this increase may be attributable to improved technical detection. 
Such rapid growth makes it even more important that a standard as widely deployed as the TCF provides reliable legal compliance.

\subsubsection{Not all apps  correctly  store user's consent banner choices.} 
\label{subsec-discussion-results-rq2}
We found 15 apps that only store user' banner choices when the user gave consent to all data processing. 
When the user withheld consent for one or more purposes, the consent banner would  reappear upon every subsequent launch of the app.
 Because the user’s choice is never registered (as demanded by the accountability principle under Article 7(1) of the GDPR),  repeated prompts can pressure users to accept out of fatigue, potentially influencing  their choices and bringing into question the freedom of consent expressed~\cite{cnil_recommendation_2025}, which ultimately may infringe a \textit{freely given consent} and thus the \textit{Lawfulness principle} (P1), the \textit{correct consent registration} requirements (LR1), and the principle of Data Protection by Design (P5).
 To this scope, such \textit{nagging} practices, commonly identified as a dark pattern, appear closely tied to  incorrect registration of consent banner choices \cite{Sant-etal-20-TechReg}. 
%

Another related practice to ``incorrect consent registration'' that occurred during dynamic analysis was inconsistent registration of banner choices across apps using the same CMP. 
Indeed, different apps implementing banners from the same CMP stored  user's choices regarding legitimate interest differently, despite the user having clicked a visually identical ``disagree to all'' button in each case. 
Besides increasing the effort required for our analysis (we had to manually opt out of every individual purpose for the \textbf{Ø}-approach), this inconsistency arguably deceives user. Users may reasonably expect that a ``disagree to all'' actually  applies to all processing purposes, not only those requiring consent. 
When this expectation is not met, a discrepancy arises between the user’s understanding and the actual legal effect of their choice.
Such practices may conflict with the \textit{Fairness principle} (P2)~\cite[para 9]{EDPB2022_DarkPatterns}. 
Consent interfaces must not mislead users about the effect of their choices.  A ``disagree to all'' button that only applies to a subset of purposes, without clearly indicating this limitation, can create a mismatch with users' reasonable expectations and the actual processing that follows. 



\subsubsection{TCF-based apps share AAID without consent at scale.} 
\label{subsec-discussion-results-rq3}
Even when users rejected consent, at least 64.3\% of the analyzed apps transmitted AAID (\textbf{Ø}-approach, in the active stage).
IAB Europe previously stated that online identifiers, such as IP addresses and AAID, constitute personal data under the GDPR~\cite{IAB-definition-personal-data}. 
Google's ad service, AdMob, declares that \textit{it is necessary to obtain consent from users in the EEA to use personal data, such as AAID}~\cite{Admob-personal-data}. 
Android documentation further describes the AAID as the recommended identifier for building profiles of users and enabling tracking. 
The same documentation also mentions that if the user  opts out of personalization using AAID -- in their Android device's settings --, AAID will be removed and made unidentifiable~\cite{Android-identifiers}. 
This implies that AAID is intended specifically for advertising-related tracking and personalization purposes.
Given that consent is the legal ground for targeted advertising (see Section~\ref{subsec:gdpr-epd}), and that AAID is a dedicated targeted advertising identifier, \textit{data controllers are required to obtain consent before  collecting it -- a position consistently reflected by stakeholders, including IAB, Google, and DPAs such as the CNIL~\cite{cnil_recommendation_2025}}.
Yet, almost 2/3 of TCF-based apps in our dataset shared AAID without consent, thereby questioning the legitimacy of the TCF to facilitate legal compliance. 
%
Our results, from both the \textit{passive} and \textit{active} stages of  traffic analysis, show  that \textbf{when TCF-based apps share users' personal data, they will in a majority of cases do so regardless of the users' banner choices}. 
The only workaround for users who do not want to share their data with apps would be to manually remove their AAID in their Android device's settings, as mentioned by Android's developers documentation~\cite{Android-identifiers}. 
This solution however relies on an opt-out solution, and therefore does not meet the \textit{Data Protection by Default principle} (P4), as it shifts the responsibility for preventing data sharing onto users rather than enforcing it by default at the application level.


\subsubsection{Game apps are the top violators} 
\label{subsec:discussion_games}
Our analysis reveals that game apps constitute the primary sources of GDPR violations.
As demonstrated in Section~\ref{subsec:games}, we found that game-related apps transmit AAID to third-parties prior to obtaining valid user consent.
This behavior shows systemic disregard for the  \textit{prior consent requirement}, thereby invalidating the legal basis for processing and directly infringing the GDPR’s \textit{Lawfulness} (P1) principle.
The implications of this practice are amplified by the demographic composition of the affected user base.
Recent industry data ~\cite{lee_mobile_2025} indicates that 48\% of mobile gamers (worldwide) are aged between 18–34, while minors aged  13-17 account for an additional 16\% of the mobile gaming user base.
These results are in line with a 2024 survey by the Oxcom (the UK Office of Communications), which states that 52\% of 16-24 years old play on mobile, and 49\% of 25-34 y.o.~\cite[p.327]{ofcom_ofcom_2024}.
Consequently, users, including children, are subjected to tracking before having any opportunity to make a meaningful choice, raising heightened concerns regarding fairness and their vulnerability.
Children's vulnerability in online gaming is currently under scrutiny, with the Irish's Garda National Cyber Crime Bureau warning that children are being ``exploited on online gaming platforms at an \textit{alarming scale}''~\cite{rte_senior_2025}.
BEUC (the European Consumer Organisation) already denounced in  2024 that children are ``are even more vulnerable to […] manipulative tactics''~\cite{beuc_consumer_2024}.
These findings are particularly timely in light of the EU’s forthcoming Digital Fairness Act (DFA), currently in draft form, which aims to curtail harmful practices in gaming ecosystems~\cite{DFA-Q-A-24}. 
\subsubsection{Google dominance}
\label{subsec:google_dominance}
The significant imbalance in CMP distribution observed — with Google LLC acting as the CMP for 89.76\% of all TCF-based apps in our dataset (see Table~\ref{tab:CMPs}) — raises concerns about \textit{platform gatekeeper power} within the Android ecosystem. 
 Google simultaneously owns Android (the Operating System), the Play Store (the distribution channel to download apps for most end-users), it is the biggest CMP for TCF-based apps by large (89.76\%), and amongst the top data broker for mobile advertising receiving personal data in our analysis (number 5, see Section~\ref{sec:domains}).
Google is even pushing further its dominant position  by requiring Android developers to register centrally~\cite{android_new_2025}, a move which triggered reactions from small players also distributing apps on Android, such as F-droid,\footnote{``In addition to demanding payment of a registration fee and agreement to their (non-negotiable and ever-changing) terms and conditions, Google will also require the uploading of personally identifying documents, including government ID, by the authors of the software[…]''~\cite{f-droid_f-droid_2025}.} before backing down~\cite{android_android_2025}.
This concentration allows Google to shape how consent is obtained and to influence which forms of data collection are facilitated or constrained at scale. 
For instance, Google claims that ``[publishers] are responsible to obtain users consent on [their] website or app or any data [they] upload to Google.''~\cite{google_consent_2025}.
However, its Android `consent mode'~\cite{mertens2025you} documentation -- intended to guide developers in configuring SDK behavior based on a user’s consent status -- provides a revealing example: it suggests setting all `consent default states' to ``true'', including those related to ad personalization signals~\cite{google_set_2025}.
In effect, research shows that developers tend to rely on default CMP configurations~\cite{utz2023privacy,toth2022dark}.
\textit{Observed violations therefore likely result from both developer choices and platform-level design.}
As a result, the compliance issues identified in Section~\ref{sec:results} cannot be attributed solely to individual app publishers: they may be structurally amplified by Google, as a platform gatekeeping actor~\cite{munir2025chrome} that sets the technical environment, provides the consent mechanism, and participates as a major data recipient. 
This suggests that effective regulatory oversight must also consider the responsibilities of dominant platform providers, and not limit accountability to app-level practices.
Google leverages control of a core platform component to reinforce its position in the advertising and tracking ecosystem, and dominates the browser market (Chrome). 
These behaviors may therefore reflect structural properties of an ecosystem organized around Google’s vertically integrated interests.

\subsubsection{Google structurally accommodates potential legal violations as a CMP}
\label{subsec:google_cmp}


Each purpose for processing personal data can rely on only one lawful basis. 
%
Because Google’s CMP solicits  consent as the chosen legal basis, this  decision effectively precludes reliance on  any other lawful basis. In other words, the app developer is prevented from invoking other legal grounds under Article 6(1) of the GDPR~\cite[para 10]{EDPB2024_legitimateinterest}\cite{noyb_news_2021} -- such as legitimate interest -- as illustrated in Figure 2b. 
This choice directly implicates the \textit{Lawfulness principle} (P1), as it constrains the controller’s ability to determine and rely upon an appropriate legal basis for processing.
%
Moreover, according to the European Data Protection Board (EDPB)~\cite[para 121-123]{edpb_guidelines_2020}, a data controller must disclose the applicable legal basis prior to  data collection, and in direct relation to each specific purpose. This entails that if an app publisher claims to rely on consent for the processing while, in practice, another lawful basis is used, this constitutes a fundamentally unfair practice toward data subjects. Consequently, the publisher cannot switch from consent to another legal basis, such as legitimate interest~\cite{morel2023legitimate}. 
As explained in Section~\ref{subsec:gdpr-epd}, advertising and related ad measurement purposes require consent. Attempting to rely on legitimate interest for advertising purposes risks violating  the \textit{Lawfulness principle} (P1)~\cite{kyi2023investigating}. 
Empirical evidence from user studies further supports this  interpretation.
Several studies show that  users are uncomfortable with the sharing of their data for the purposes of ``personalized ad delivery and measurement’’ and ``personalized content delivery and measurement’’~\cite{ChancharyChiasson2015, KozyrevaLorenzSpreenHertwigLewandowskyHerzog2021, kyi2023investigating}. Furthermore, users disapprove of websites that list multiple advertising purposes under the legal basis of legitimate interest, a practice that is not compliant with data protection law~\cite[para 118]{CJEU_C-252-21_Meta_v_Bundeskartellamt_2023}\cite{KyiMhaidliSantosRoesnerBiega2024}. 
Given that Google’s CMP relies on legitimate interest in some contexts while simultaneously defaulting consent to ``accept'' for ad-related purposes, it is arguable that Google’s CMP may process personal data for its own purposes. 
Under such circumstances, Google’s role could be interpreted as that of a joint controller alongside app developers, thereby requiring it to comply with the full set of GDPR obligations.
The rate of potential violations observed in other CMPs (Easybrain Ltd and Outfit7 Limited, 100\% of apps caught sharing AAID without consent in both stages of our analysis, see Section~\ref{subsec:per_dev}) also supports the consideration of a \textit{more careful examination of the role of CMPs in mobile contexts}.

%



\subsubsection{Bad defaults used by the CMP Google LLC}

As exemplified in Figure~\ref{fig:cmp-300-options} of Section~\ref{subsec:auto_interaction}, 
the purpose of \textit{measuring advertising performance} shows the legal basis of consent deselected and legitimate interest selected for 45 vendors.
In practice, all banners in our dataset provided by CMP 300 (Google) preselected all legitimate interest purposes for all vendors.
When data settings are preselected, users are subject to a specific data protection level, determined by
the provider by default, rather than by users.  
The ``salient field''  of the consent banner provides the most important information to aid the user's decision.  
Making some buttons more salient is an example of design outcomes that are intended to surreptitiously nudge users by making a pre-chosen and intended choice more salient~\cite[pp. 19-20]{NCC-Deceived-18}. 
As the most data invasive option is enabled by default, this constitutes the dark
pattern of \textit{preselection or bad defaults}~\cite{EDPB2022_DarkPatterns, Gray2021DarkPatternsConsent}. 
Bad defaults subvert user’s reasonable expectation that default settings will be in their best interest, instead requiring users to take
active steps to change settings that may cause harm or unintentional disclosure of personal data~\cite{GraySantosBielovaMildner2024}.
The practice of preselection is prohibited  by the Court of Justice of the EU~\cite[para 57, 63]{CJEU_Planet49_2019}, and by the EDPB~\cite[para 54] {EDPB2022_DarkPatterns}. Such practices may also violate the \textit{Data Protection by Default principle} (P4), as they shift the burden of privacy protection from the controller to the user.

\subsubsection{Potential US personal data transfers without required safeguards}
\label{subsec:US_transfer}
The results show in Section~\ref{subsec:cat_count} that 82.14\% of TCF-based apps are developed outside the EU/EEA. 
Because the GDPR applies extraterritorially to  actors processing personal data of EU users (Article 3), these app publishers are still subject to its requirements. 
Moreover, when personal data is transmitted to a third-country, the GDPR’s international data transfer rules are triggered.
Apps send personal data to ad-tech servers that belong to companies headquartered in the US (see Section~\ref{sec:domains}). 
While we cannot conclusively determine the physical location of the servers involved—given the well -- known difficulty of geolocating infrastructure behind domains~\cite{laouar2025rethinking} and the widespread use of Content Delivery Networks (CDNs) -- we cannot exclude the possibility that personal data is transferred to the US. This uncertainty is further compounded by the possibility of subsequent ex-post transfers for processing.
%
Under EU data protection law, transfers of personal data to the US are restricted due to the risk of US intelligence access.
EU-US data transfers require explicit consent, contractual, organizational, and technical measures to prevent access by US intelligence services. 
Consequently, the transmission of unique identifiers and IP addresses to US-based entities may raise compliance concerns where no valid transfer mechanism or adequate safeguards are in place. In such circumstances, the level of protection required under Article 44 GDPR may be undermined, potentially implicating the \textit{Lawfulness} (P1) and \textit{Security} (P6) principles.

\subsection{Limitations}
\label{subsec:limitations}

\subsubsection{Dataset limitations}
Our dataset comprises several limitations, our results should therefore be interpreted in this light.
First, we limited our collection to top apps per category, which excluded less popular apps.
Second, 585 (11.5\%) of apps failed to download for various reasons (such as the detection of an emulated device, see Section~\ref{sec:result-rq1}).
Third, we only studied a subset of the CMPs available on mobile.
The TCF lists 50 different CMPs on mobile (exclusive, and combined web/mobile) at the time of this study~\cite{TCF-CMP-list}, these CMPs provide  visually different banners and therefore require different heuristics to interact with.
We captured 12 different CMPs amongst apps initializing the framework, and we interacted with six out of the 12. 
Each of the six CMPs we did not interact with either only appeared once in unique apps (5/6), or had a unique design in each app (1/6), we therefore discarded their analysis (only 1.90\% of apps were discarded for this reason in this step of the analysis).
Our data analysis is therefore biased towards apps relying on standardized CMPs, and the potential violations detected may not be fully representative of the broader Android ecosystem.



\subsubsection{Capturing transformed data}
\label{sec:transformed}
Although we filtered a range of personal data (see Section~\ref{sec:workflow}), our analysis does not account for \textit{transformed data}, \textit{i.e.}, hashed (SHA) or encoded (Base64) data.
A manual investigation into an arbitrary sample of decrypted packets did not reveal transformed data, we thus proceeded by not performing additional analysis.
However, we acknowledge that an extended analysis may have captured other types of personal data.
Since, the list of data considered is static and relatively small, a pre-computation of the hashes and encodings can bring the evaluation to a reasonable cost.


\subsubsection{Network traffic on emulated devices}
\label{subsec:emu}
The reliance on emulated devices, instead of physical devices~\cite{koch2023ok, nguyen2021share}, may have altered the network behavior of analyzed apps.
This design choice however enabled us to streamline the analysis (see Section~\ref{subsec:snapshots}), and we contend that the risk of having our results compromised is limited.
To the best of our knowledge, a systematic comparison between traffic generated by physical and emulated devices has not yet been reported. However, emulators are commonly adopted in prior traffic analysis studies as a practical, reproducible, and accepted experimental platform.
Surveys and papers on mobile traffic analysis and tracking do not mention it as a limitation~\cite{binns_tracking_2022, jin_why_2018, binns_measuring_2018}.
Although Pegioudis et al.~\cite{pegioudis_not_2023} warn that it could alter behavior in other contexts (mobile browser traffic) since websites could detect such a ``non-realistic'' visit, we argue that our technical setting is incomparable: websites are third-parties, while apps are first-parties in our context.
In a widely cited survey on mobile network traffic analysis, Conti et al.~\cite{conti_dark_2018} actually present emulated devices are a valid alternative to palliate the difficulty of automating tests: ``it constitutes an ideal point of capturing for mobile traffic. This approach is particularly useful if the focus of the analysis is on the network traffic of a specific mobile app''.

\subsubsection{Analyzing other operating Systems} 
\label{subsec:ios}
The methodology used in this paper is dependent on \verb|adb| which is unique for Android devices, making extensions of our methodology to other operating systems complicated. 
Furthermore, iOS apps ask for users' permission to use advertising IDs~\cite{kollnig_goodbye_2022} (opt-in basis, unlike AAID), making it less likely for apps to transmit user data unknowingly~\cite{apple-idfa}. 
Note also that Android is the most widely used operating system~\cite{mobile-os-marketshare}.
Lastly, only focusing on Android allows for a more in-depth analysis.
Indeed, Apple prevents the rooting of its devices, making a MITM approach difficult, only leaving as a valid approach the tampering with a certificate to later decrypt the content of the network traffic.
Recent work on iOS still relies on metadata analysis~\cite{moti2025whispertest}.

\subsection{Future Work}
\label{subsec:future}


\subsubsection{Extending the dataset}
Assessing the presence of the TCF on an app required downloading it, amounting to a significant overhead (87.15\% of downloaded apps did not use the TCF).
During interactions with consent banners, we noticed that many apps using the TCF were published by the same developers. 
In the Google Play Store, each application has a link to its developer,\footnote{The Play Store refers to the developer, although the legal responsibility is first on the publisher~\cite{cnil_recommendation_2025}.} meaning that, given an app, a simple heuristic consisting in scraping all apps from the same developer can rapidly expand our dataset of TCF-based apps.
We tried this method on a set of 842 applications initializing the framework, and found 4741 apps initializing the TCF too, all published by developers using the TCF in other apps. 
This method can bootstrap the creation of a larger dataset, enabling further privacy analysis, 
with the downside of not providing further data on proportions of TCF usage in the Play Store, and possibly biasing the representativeness of results.
Another possibility of extension is to use a different sampling method, by downloading not only top apps but also long tail apps with fewer downloads.


\subsubsection{Further interaction with apps} 
Our traffic analysis was only conducted on apps while interacting with their consent banners, and then while idling afterward. 
Our automated interactions may therefore  not have accurately captured realistic human-like interactions.
This leaves most of the applications' functionalities untouched, potentially missing out sharing of personal data.
Further interaction with apps and additional collection points may yield better insights into data collection in this context.
For instance, data could be collected before any interaction, or after realistic simulation with an app's features.
However, automating interactions with apps is challenging, since apps are unique.
Additionally, how users interact with consent banners on mobile has only been little studied~\cite{jha2023refuse}, which forces us to formulate hypotheses in the absence of literature on the topic (\textit{i.e.}, that users behave similarly than with web consent banners~\cite{kyi2023investigating}).
A longitudinal study~\cite{wijesekera2017feasibility} with real participants over a longer period of time may address this challenge, and provide a different picture.
The combination of realistic interactions and analysis of transformed data (see Section~\ref{sec:transformed}) frames an interesting yet challenging research avenue.

\subsubsection{Comparing emulated and physical devices}
Another promising research avenue is the comparison between emulated and physical devices.
As discussed in Section~\ref{subsec:emu}, the use of emulators is considered standard practice for network traffic analysis.
However, there is no strict guarantee that this aspect of our setup did not impact our results.
Therefore, future work should encompass both emulated and physical devices to provide a clearer picture.


\subsubsection{The state of affairs on iOS}
As mentioned in Section~\ref{subsec:ios}, the present paper only focuses on the analysis of Android apps.
Our results cover issues related to Google/Alphabet's dominant position (as owner of the platform distributing apps, the OS, main CMP, and data broker), which visibly allows legal violations at scale.
While the  main mobile OS provider, Apple, historically relied on its image as a privacy-conscious company~\cite{vincent_apple_2016,zuboff2023age}, a claim not always met in reality~\cite{roberts-islam_siri_2025}.
A natural and promising research avenue therefore is the study of the TCF on iOS, to systematically assess whether apps distributed through the App Store (Apple's equivalent of the Play Store) genuinely respect users' consent choices.



\section{Conclusion}
\label{section-conclusion}


This paper investigates the prevalence and the usage of the TCF in Android applications. 
To quantify the usage of the TCF in Android apps, we scraped and downloaded 4482 of the most popular apps in the Google Play Store.
We found that \textbf{576 apps were identified as implementing the TCF, representing 12.85\%} of the downloaded apps. 
This indicates an increase in usage of the framework compared to the closest related work~\cite{altpeter2022ic,koch2023ok}.
The TCF can be found in apps of all categories, developed in various countries including outside the EU/EEA, with Google being by far the main CMP in our dataset (89.76\%).
We automatically interacted with consent banners of 576 applications and confirmed their stored banner choices. 
As a result, we found that \textbf{15 applications only stored our choices if provided with consent to all data processing}, 
suggesting that certain TCF-based apps do not respect users’ banner choices.
We analyzed network traffic in two stages: passive (post-choices) and active (during interaction with the banner/before choices were enacted).
Our results show that \textbf{66.2\% of applications transmitted AAID}, a unique identifier dedicated to targeted advertising, when disagreeing to all in the \textbf{passive}-stage. 
While \textbf{55.3\% of apps also transmitted AAID while we were interacting with the apps' banners} (before choices were made) during the \textbf{active}-stage of our analysis.
Games are the top app categories sharing AAID without consent.
We conclude that the TCF fails to facilitate legal compliance at scale on Android apps. 
Although IAB Europe announced in June 2025 a new minor version of the TCF (v.2.3)~\cite{iab_europe_release_2025} (see Section~\ref{subsec:TCF}), it is unlikely that the modifications will address the issues identified in this paper~\cite{usercentrics_iab_2025}.



\begin{acks}
This work was partially supported by the Wallenberg AI, Autonomous Systems and Software Program (WASP) funded by the Knut and Alice Wallenberg Foundation, and by the Utrecht Centre for Regulation and Enforcement in Europe (RENFORCE).
The authors would like to thank the reviewers for helping us improve this paper. 
\end{acks}


\bibliographystyle{ACM-Reference-Format}
\bibliography{sample-base}

\appendix
\section{Ethical Considerations}
\label{sec:ethics}
Our compliance assessment of TCF-based apps is constrained by the availability of apps in our dataset, which was shaped by ethical and technical boundaries. 
A critical consideration was adherence to the Play Store’s \verb|robots.txt|~\cite{play-robot.txt}. 
This file defines permissible and prohibited interactions for automated systems~\cite{Cloudflare-robots}, ensuring respect for the platform’s terms of use.
Notably, the \verb|robots.txt| restrictions explicitly prohibit non-human users, such as web scrapers, from utilizing the Play Store search bar. 
However, it permitted access to top apps by category, as this was not restricted. 
To maintain ethical compliance, we limited our dataset to apps accessible through these means, prioritizing respect for platform guidelines over broader data collection.


Prior to publication, we performed a responsible disclosure to developers of apps sharing AAID without consent.
We also plan to share our results with the European Data Protection Board, data protection regulators, as well as with IAB Europe.


\section{Tables}

\begin{table}[tph]
\caption{App categories from our dataset and their usage of the TCF}
\label{appendix:categories}

\centering
\begin{tabular}{|p{0.45\linewidth}|p{0.45\linewidth}|}
\hline
\textbf{Category} & \textbf{Usage \%} \\
\hline
PERSONALIZATION & 54/101 (53.47\%) \\
\hline
LIBRARIES\_AND\_DEMO & 26/76 (34.21\%) \\
\hline
ART\_AND\_DESIGN & 27/79 (34.18\%) \\
\hline
VIDEO\_PLAYERS & 20/67 (29.85\%) \\
\hline
TOOLS & 26/90 (28.89\%) \\
\hline
PHOTOGRAPHY & 22/77 (28.57\%) \\
\hline
MAPS\_AND\_NAVIGATION & 22/86 (25.58\%) \\
\hline
PRODUCTIVITY & 16/67 (23.88\%) \\
\hline
MUSIC\_AND\_AUDIO & 20/85 (23.53\%) \\
\hline
WEATHER & 15/65 (23.08\%) \\
\hline
FAMILY\_EDUCATION & 2/11 (18.18\%) \\
\hline
GAME\_BOARD & 13/92 (14.13\%) \\
\hline
COMMUNICATION & 8/57 (14.04\%) \\
\hline
GAME\_WORD & 13/93 (13.98\%) \\
\hline
GAME\_TRIVIA & 9/66 (13.64\%) \\
\hline
BEAUTY & 7/52 (13.46\%) \\
\hline
HEALTH\_AND\_FITNESS & 12/91 (13.19\%) \\
\hline
GAME\_ARCADE & 12/93 (12.9\%) \\
\hline
EVENTS & 11/86 (12.79\%) \\
\hline
GAME\_MUSIC & 10/85 (11.76\%) \\
\hline
GAME\_PUZZLE & 13/113 (11.5\%) \\
\hline
PARENTING & 7/62 (11.29\%) \\
\hline
ENTERTAINMENT & 8/75 (10.67\%) \\
\hline
FAMILY & 9/86 (10.47\%) \\
\hline
GAME\_SIMULATION & 12/115 (10.43\%) \\
\hline
GAME\_SPORTS & 11/107 (10.28\%) \\
\hline
DATING & 6/60 (10.0\%) \\
\hline
NEWS\_AND\_MAGAZINES & 6/61 (9.84\%) \\
\hline
SOCIAL & 7/80 (8.75\%) \\
\hline
GAME\_CARD & 6/75 (8.0\%) \\
\hline
HOUSE\_AND\_HOME & 5/64 (7.81\%) \\
\hline
BOOKS\_AND\_REFERENCE & 6/77 (7.79\%) \\
\hline
GAME\_CASUAL & 10/130 (7.69\%) \\
\hline
GAME\_RACING & 8/105 (7.62\%) \\
\hline
GAME\_ADVENTURE & 8/108 (7.41\%) \\
\hline
GAME\_ACTION & 12/162 (7.41\%) \\
\hline
GAME\_ROLE\_PLAYING & 8/119 (6.72\%) \\
\hline
GAME\_EDUCATIONAL & 10/156 (6.41\%) \\
\hline
LIFESTYLE & 5/80 (6.25\%) \\
\hline
GAME\_CASINO & 4/66 (6.06\%) \\
\hline
COMICS & 3/53 (5.66\%) \\
\hline
GAME & 26/484 (5.37\%) \\
\hline
FAMILY\_CREATE & 2/40 (5.0\%) \\
\hline
SPORTS & 5/100 (5.0\%) \\
\hline
BUSINESS & 4/82 (4.88\%) \\
\hline
GAME\_STRATEGY & 4/98 (4.08\%) \\
\hline
AUTO\_AND\_VEHICLES & 2/51 (3.92\%) \\
\hline
\end{tabular}
\end{table}

\begin{table}[t]
\centering
\begin{tabular}{|p{0.45\linewidth}|p{0.45\linewidth}|}
\hline
EDUCATION & 3/82 (3.66\%) \\
\hline
TRAVEL\_AND\_LOCAL & 3/94 (3.19\%) \\
\hline
MEDICAL & 1/32 (3.12\%) \\
\hline
APPLICATION & 4/144 (2.78\%) \\
\hline
FAMILY\_MUSICVIDEO & 1/44 (2.27\%) \\
\hline
FOOD\_AND\_DRINK & 1/74 (1.35\%) \\
\hline
SHOPPING & 0/62 (0\%) \\
\hline
FAMILY\_BRAINGAMES & 0/66 (0\%) \\
\hline
FINANCE & 0/71 (0\%) \\
\hline
FAMILY\_PRETEND & 0/16 (0\%) \\
\hline
FAMILY\_ACTION & 0/54 (0\%) \\
\hline
Data Unavailable & 0/11 (0\%) \\
\hline
\end{tabular} 
\vspace{5mm}
\end{table}


\begin{table}[tph]
\caption{Distribution of countries where TCF-based apps in our dataset were developed}

\label{appendix:developer-country-statistics}
\centering
\begin{tabular}{|p{0.45\linewidth}|p{0.45\linewidth}|}
\hline
\textbf{Country} & \textbf{Apps using the TCF (N = 576)}\\
\hline
Vietnam & 94 (16.32\%) \\
\hline
Hong Kong & 60 (10.42\%) \\
\hline
Cyprus & 41 (7.12\%) \\
\hline
China & 41 (7.12\%) \\
\hline
Israel & 36 (6.25\%) \\
\hline
India & 36 (6.25\%) \\
\hline
App Removed & 28 (4.86\%) \\
\hline
United States & 27 (4.69\%) \\
\hline
Pakistan & 22 (3.82\%) \\
\hline
United Kingdom & 17 (2.95\%) \\
\hline
United Arab Emirates & 15 (2.6\%) \\
\hline
Sweden & 14 (2.43\%) \\
\hline
Morocco & 14 (2.43\%) \\
\hline
Data Unavailable & 11 (1.91\%) \\
\hline
Singapore & 11 (1.91\%) \\
\hline
South Korea & 10 (1.74\%) \\
\hline
Türkiye & 10 (1.74\%) \\
\hline
Spain & 10 (1.74\%) \\
\hline
Japan & 9 (1.56\%) \\
\hline
Andorra & 5 (0.87\%) \\
\hline
Poland & 5 (0.87\%) \\
\hline
Canada & 5 (0.87\%) \\
\hline
Germany & 5 (0.87\%) \\
\hline
Romania & 4 (0.69\%) \\
\hline
Moldova & 4 (0.69\%) \\
\hline
Serbia & 3 (0.52\%) \\
\hline
Brazil & 3 (0.52\%) \\
\hline
Hungary & 3 (0.52\%) \\
\hline
Portugal & 3 (0.52\%) \\
\hline
Italy & 2 (0.35\%) \\
\hline
Czechia & 2 (0.35\%) \\
\hline
Russia & 2 (0.35\%) \\
\hline
Mexico & 2 (0.35\%) \\
\hline
Cayman Islands & 2 (0.35\%) \\
\hline
British Virgin Islands & 2 (0.35\%) \\
\hline

\end{tabular}
\end{table}

\begin{table}[tph]
\centering
\begin{tabular}{|p{0.45\linewidth}|p{0.45\linewidth}|}
\hline
Australia & 2 (0.35\%) \\
\hline
Lithuania & 1 (0.17\%) \\
\hline
Netherlands & 1 (0.17\%) \\
\hline
Malaysia & 1 (0.17\%) \\
\hline
France & 1 (0.17\%) \\
\hline
Estonia & 1 (0.17\%) \\
\hline
Ukraine & 1 (0.17\%) \\
\hline
Switzerland & 1 (0.17\%) \\
\hline
Malta & 1 (0.17\%) \\
\hline
Yemen & 1 (0.17\%) \\
\hline
Azerbaijan & 1 (0.17\%) \\
\hline
New Zealand & 1 (0.17\%) \\
\hline
Bangladesh & 1 (0.17\%) \\
\hline
Indonesia & 1 (0.17\%) \\
\hline
Marshall Islands & 1 (0.17\%) \\
\hline
Uruguay & 1 (0.17\%) \\
\hline
Norway & 1 (0.17\%) \\
\hline
\end{tabular} 
\vspace{5mm}
\end{table}

\begin{table}[tph]
\caption{Apps incorrectly registering banner choices.}
\label{tab:incorrec-registration}
\begin{tabular}{|l|}
\hline
App names and packages \\ \hline
\begin{tabular}[c]{@{}l@{}}Sweet Beauty Camera Filter\\ com.camera.sweet.beauty.fillter.cameraeditor\end{tabular} \\ \hline
\begin{tabular}[c]{@{}l@{}}Geotag Photo: Camera Location\\ com.cameratag.geotagphoto.gpscamera\end{tabular} \\ \hline
\begin{tabular}[c]{@{}l@{}}Control Center Simple\\ com.tools.control.center.simplecontrol\end{tabular} \\ \hline
\begin{tabular}[c]{@{}l@{}}Draw Animation - Anim Creator\\ com.banix.drawsketch.animationmaker\end{tabular} \\ \hline
\begin{tabular}[c]{@{}l@{}}Silly Smiles Live Wallpaper 4K\\ com.silly.smilewallpaper.sillywallpaper\end{tabular} \\ \hline
\begin{tabular}[c]{@{}l@{}}AR Draw Sketch: Trace \& Paint\\ com.ardrawing.sketch.trace.paint.drawsketch.aipainting.paper\end{tabular} \\ \hline
\begin{tabular}[c]{@{}l@{}}Easy Piano Keyboard - Piano88\\ com.banix.piano.learnpiano\end{tabular} \\ \hline
\begin{tabular}[c]{@{}l@{}}Find Phone by Clap, Whistle\\ com.bigsoft.clap.myphone\end{tabular} \\ \hline
\begin{tabular}[c]{@{}l@{}}Caller ID Name, Location \& SMS\\ com.indiastudio.caller.truephone\end{tabular} \\ \hline
\begin{tabular}[c]{@{}l@{}}Chat AI - AI Chatbot Friend\\ com.antonio.chat.ai.free.artificial.intelligence\end{tabular} \\ \hline
\begin{tabular}[c]{@{}l@{}}Dating - Chat \& Meet Singles\\ com.appgame.dating.app.free.chat.flirt.dating\end{tabular} \\ \hline
\begin{tabular}[c]{@{}l@{}}Locate Mobile by Number\\ com.appgame.phone.number.locator.by.number\end{tabular} \\ \hline
\begin{tabular}[c]{@{}l@{}}Phone Tracker By Number\\ com.antonio.family.locator.phone.tracker\end{tabular} \\ \hline
\begin{tabular}[c]{@{}l@{}}Books - Read and Download\\ com.appgame.free.books.read.free.books\end{tabular} \\ \hline
\begin{tabular}[c]{@{}l@{}}PDF Viewer: PDF Fill \& Sign\\ com.fillpdf.pdfeditor.pdfsign\end{tabular} \\ \hline
\end{tabular}
\end{table}

\begin{table}[t]
\caption{The share of TCF-based applications in each category from our dataset (combined passive and active stages, on both \textbf{LI}- and \textbf{Ø}-approaches), that transmitted AAID}
\label{tab:cross_categories_sharing}

\centering
\begin{tabular}{|p{0.68\linewidth}|p{0.22\linewidth}|}
\hline
\textbf{Category} & \textbf{Violation \%} \\
\hline
GAME\_SIMULATION & 12/12 (100.0\%) \\
\hline
GAME\_ARCADE & 12/12 (100.0\%) \\
\hline
GAME\_SPORTS & 11/11 (100.0\%) \\
\hline
GAME\_CASUAL & 10/10 (100.0\%) \\
\hline
GAME\_CASINO & 4/4 (100.0\%) \\
\hline
EDUCATION & 3/3 (100.0\%) \\
\hline
AUTO\_AND\_VEHICLES & 2/2 (100.0\%) \\
\hline
GAME\_BOARD & 12/13 (92.31\%) \\
\hline
GAME & 23/26 (88.46\%) \\
\hline
PRODUCTIVITY & 14/16 (87.5\%) \\
\hline
GAME\_ADVENTURE & 7/8 (87.5\%) \\
\hline
GAME\_ROLE\_PLAYING & 7/8 (87.5\%) \\
\hline
ENTERTAINMENT & 7/8 (87.5\%) \\
\hline
MUSIC\_AND\_AUDIO & 17/20 (85.0\%) \\
\hline
GAME\_PUZZLE & 11/13 (84.62\%) \\
\hline
GAME\_WORD & 11/13 (84.62\%) \\
\hline
GAME\_ACTION & 10/12 (83.33\%) \\
\hline
NEWS\_AND\_MAGAZINES & 5/6 (83.33\%) \\
\hline
DATING & 5/6 (83.33\%) \\
\hline
GAME\_CARD & 5/6 (83.33\%) \\
\hline
GAME\_MUSIC & 8/10 (80.0\%) \\
\hline
SPORTS & 4/5 (80.0\%) \\
\hline
HEALTH\_AND\_FITNESS & 9/12 (75.0\%) \\
\hline
GAME\_RACING & 6/8 (75.0\%) \\
\hline
WEATHER & 11/15 (73.33\%) \\
\hline
TOOLS & 19/26 (73.08\%) \\
\hline
MAPS\_AND\_NAVIGATION & 16/22 (72.73\%) \\
\hline
SOCIAL & 5/7 (71.43\%) \\
\hline
ART\_AND\_DESIGN & 19/27 (70.37\%) \\
\hline
VIDEO\_PLAYERS & 14/20 (70.0\%) \\
\hline
PHOTOGRAPHY & 15/22 (68.18\%) \\
\hline
GAME\_TRIVIA & 6/9 (66.67\%) \\
\hline
BOOKS\_AND\_REFERENCE & 4/6 (66.67\%) \\
\hline
PERSONALIZATION & 35/54 (64.81\%) \\
\hline
LIFESTYLE & 3/5 (60.0\%) \\
\hline
APPLICATION & 2/4 (50.0\%) \\
\hline
HOUSE\_AND\_HOME & 2/5 (40.0\%) \\
\hline
COMMUNICATION & 3/8 (37.5\%) \\
\hline
EVENTS & 4/11 (36.36\%) \\
\hline
COMICS & 1/3 (33.33\%) \\
\hline
TRAVEL\_AND\_LOCAL & 1/3 (33.33\%) \\
\hline
LIBRARIES\_AND\_DEMO & 8/26 (30.77\%) \\
\hline
PARENTING & 2/7 (28.57\%) \\
\hline
BEAUTY & 2/7 (28.57\%) \\
\hline
GAME\_STRATEGY & 1/4 (25.0\%) \\
\hline
GAME\_EDUCATIONAL & 1/10 (10.0\%) \\
\hline
Data Unavailable & 0/11 (0\%) \\
\hline
FAMILY & 0/9 (0\%) \\
\hline
BUSINESS & 0/4 (0\%) \\
\hline
FAMILY\_CREATE & 0/2 (0\%) \\
\hline
FAMILY\_EDUCATION & 0/2 (0\%) \\
\hline
FAMILY\_MUSICVIDEO & 0/1 (0\%) \\
\hline
FOOD\_AND\_DRINK & 0/1 (0\%) \\
\hline
MEDICAL & 0/1 (0\%) \\
\hline
\end{tabular} 
\vspace{5mm}
\end{table}

\begin{table}[t]
\caption{Developer passive stage data}
\label{tab:violating_developers_stage_1}

\centering
\begin{tabular}{|p{0.68\linewidth}|p{0.22\linewidth}|}
\hline
\textbf{Developer (name/email)} & \textbf{Violation \%} \\
\hline
Easybrain & 14/14 (100.0\%) \\
\hline
Outfit7 Limited & 9/9 (100.0\%) \\
\hline
BoomBit Games & 7/7 (100.0\%) \\
\hline
CrazyLabs LTD & 5/5 (100.0\%) \\
\hline
Battery Stats Saver & 5/5 (100.0\%) \\
\hline
gameone & 4/4 (100.0\%) \\
\hline
RobTop Games & 4/4 (100.0\%) \\
\hline
nguyenluan@lutech.ltd & 4/4 (100.0\%) \\
\hline
hoahuongduong170120@gmail.com & 4/4 (100.0\%) \\
\hline
SayGames Ltd & 3/3 (100.0\%) \\
\hline
Coco Play By TabTale & 3/3 (100.0\%) \\
\hline
Prometheus Interactive LLC & 3/3 (100.0\%) \\
\hline
LoveColoring Game & 3/3 (100.0\%) \\
\hline
KAYAC Inc. & 3/3 (100.0\%) \\
\hline
ZenLife Games Ltd & 3/3 (100.0\%) \\
\hline
ThemeKit & 3/3 (100.0\%) \\
\hline
trusted.mobile.app@gmail.com & 3/3 (100.0\%) \\
\hline
IVYGAMES & 3/3 (100.0\%) \\
\hline
FMPROJECT LIMITED & 2/2 (100.0\%) \\
\hline
LifeSim & 2/2 (100.0\%) \\
\hline
KamaGames & 2/2 (100.0\%) \\
\hline
publishing.apero@gmail.com & 2/2 (100.0\%) \\
\hline
LifePulse Puzzle Game Studio & 2/2 (100.0\%) \\
\hline
Gameberry Labs & 2/2 (100.0\%) \\
\hline
GAME OFFLINE HAY & 2/2 (100.0\%) \\
\hline
Hippo Lab & 2/2 (100.0\%) \\
\hline
MAGIC SEVEN CO., LIMITED & 2/2 (100.0\%) \\
\hline
HDuo Fun Games & 2/2 (100.0\%) \\
\hline
Maxlabs Photo Editor & 2/2 (100.0\%) \\
\hline
Color App Team & 2/2 (100.0\%) \\
\hline
FunGear inc & 2/2 (100.0\%) \\
\hline
Eyewind & 2/2 (100.0\%) \\
\hline
Braly JSC & 2/2 (100.0\%) \\
\hline
vinhdn07@gmail.com & 2/2 (100.0\%) \\
\hline
admin@vtn.global & 2/2 (100.0\%) \\
\hline
SUPERIOR  STUDIO & 2/2 (100.0\%) \\
\hline
Appwest Limited & 2/2 (100.0\%) \\
\hline
Metaverse Labs & 2/2 (100.0\%) \\
\hline
Beat Blend Labs & 2/2 (100.0\%) \\
\hline
Red Sky Labs & 2/2 (100.0\%) \\
\hline
Solitaire Aquarium & 2/2 (100.0\%) \\
\hline
Mortys Games & 2/2 (100.0\%) \\
\hline
HK Hero Entertainment Co., Limited & 2/2 (100.0\%) \\
\hline
Adkins Studio & 2/2 (100.0\%) \\
\hline
Dainik Bhaskar Group & 2/2 (100.0\%) \\
\hline
Gianluca Cisana & 1/1 (100.0\%) \\
\hline
DIC-o & 1/1 (100.0\%) \\
\hline
mobileapps@tubema.ltd & 1/1 (100.0\%) \\
\hline
Rainberry, Inc. & 1/1 (100.0\%) \\
\hline
ORANGE GAME & 1/1 (100.0\%) \\
\hline
\end{tabular}
\end{table}

\begin{table}[t]
\centering
\begin{tabular}{|p{0.68\linewidth}|p{0.22\linewidth}|}
\hline
Fearless Games sp. j. & 1/1 (100.0\%) \\
\hline
TabTale & 1/1 (100.0\%) \\
\hline
Escape Adventure Games & 1/1 (100.0\%) \\
\hline
WONDER GROUP & 1/1 (100.0\%) \\
\hline
Estoty & 1/1 (100.0\%) \\
\hline
Supercent, Inc. & 1/1 (100.0\%) \\
\hline
cerdillac & 1/1 (100.0\%) \\
\hline
mulanidhvani@gmail.com & 1/1 (100.0\%) \\
\hline
FALCON GAME & 1/1 (100.0\%) \\
\hline
TFive  Labs & 1/1 (100.0\%) \\
\hline
changpeng & 1/1 (100.0\%) \\
\hline
Dream Tap & 1/1 (100.0\%) \\
\hline
OUTLOU:D GAMES & 1/1 (100.0\%) \\
\hline
Better World Games & 1/1 (100.0\%) \\
\hline
Nebuchadnezzar DOO & 1/1 (100.0\%) \\
\hline
KTW Apps & 1/1 (100.0\%) \\
\hline
Free VPN Planet & 1/1 (100.0\%) \\
\hline
Digipom & 1/1 (100.0\%) \\
\hline
Audify Player. & 1/1 (100.0\%) \\
\hline
Colorful Point & 1/1 (100.0\%) \\
\hline
dammuhzaj@gmail.com & 1/1 (100.0\%) \\
\hline
Siti Berkah Studio & 1/1 (100.0\%) \\
\hline
Diavostar PTE. LTD & 1/1 (100.0\%) \\
\hline
developer@vottak.app & 1/1 (100.0\%) \\
\hline
ASD Dev Video Player for All Format & 1/1 (100.0\%) \\
\hline
Lotum one GmbH & 1/1 (100.0\%) \\
\hline
Clappy LLC & 1/1 (100.0\%) \\
\hline
Qiiwi Games AB & 1/1 (100.0\%) \\
\hline
Andrey Solovyev & 1/1 (100.0\%) \\
\hline
Qbis Studio & 1/1 (100.0\%) \\
\hline
Buddies Games Inc. & 1/1 (100.0\%) \\
\hline
Loop Games & 1/1 (100.0\%) \\
\hline
ZenTint Creations & 1/1 (100.0\%) \\
\hline
7788`s & 1/1 (100.0\%) \\
\hline
MOUNTAIN GAME & 1/1 (100.0\%) \\
\hline
Doodle Mobile Ltd. & 1/1 (100.0\%) \\
\hline
CanaryDroid & 1/1 (100.0\%) \\
\hline
Macro Tap & 1/1 (100.0\%) \\
\hline
1SOFT & 1/1 (100.0\%) \\
\hline
Emily Yu & 1/1 (100.0\%) \\
\hline
Prop studio & 1/1 (100.0\%) \\
\hline
haphuong4251@gmail.com & 1/1 (100.0\%) \\
\hline
Fancolor & 1/1 (100.0\%) \\
\hline
Miniclip.com & 1/1 (100.0\%) \\
\hline
CASUAL AZUR GAMES & 1/1 (100.0\%) \\
\hline
Theago Liddell & 1/1 (100.0\%) \\
\hline
Warrior Game & 1/1 (100.0\%) \\
\hline
Pixel Art - draw in fun & 1/1 (100.0\%) \\
\hline
AppQuantum & 1/1 (100.0\%) \\
\hline
EasyFun Puzzle Game Studio & 1/1 (100.0\%) \\
\hline
\end{tabular}
\end{table}

\begin{table}[t]
\centering
\begin{tabular}{|p{0.68\linewidth}|p{0.22\linewidth}|}
\hline
ZiMAD & 1/1 (100.0\%) \\
\hline
RED BRIX COMPUTER SYSTEMS & 1/1 (100.0\%) \\
\hline
Focus apps & 1/1 (100.0\%) \\
\hline
Kitten Doll HK & 1/1 (100.0\%) \\
\hline
Kitten doll & 1/1 (100.0\%) \\
\hline
Meeto Team & 1/1 (100.0\%) \\
\hline
BUZZMEDIA INC. & 1/1 (100.0\%) \\
\hline
Smart AI DEV & 1/1 (100.0\%) \\
\hline
checksmartapp@gmail.com & 1/1 (100.0\%) \\
\hline
cghxstudio@gmail.com & 1/1 (100.0\%) \\
\hline
sunflower studio & 1/1 (100.0\%) \\
\hline
aimirror-support@polyversestudio.com & 1/1 (100.0\%) \\
\hline
Outdoing Apps & 1/1 (100.0\%) \\
\hline
Pixl Concerto Tech & 1/1 (100.0\%) \\
\hline
ITO Technologies, Inc. & 1/1 (100.0\%) \\
\hline
o16i Apps & 1/1 (100.0\%) \\
\hline
Clever Apps Pte. Ltd. & 1/1 (100.0\%) \\
\hline
Coreup Teknoloji Limited Şirketi & 1/1 (100.0\%) \\
\hline
Scopely & 1/1 (100.0\%) \\
\hline
uschultz & 1/1 (100.0\%) \\
\hline
Funny Design Keyboard Themes & 1/1 (100.0\%) \\
\hline
Delicate theme for Android App & 1/1 (100.0\%) \\
\hline
Esame Marketing Limited & 1/1 (100.0\%) \\
\hline
Fancy Studio Mods & 1/1 (100.0\%) \\
\hline
Live Wallpapers and Emoji Keyboard Themes & 1/1 (100.0\%) \\
\hline
App Soft Studio & 1/1 (100.0\%) \\
\hline
chudang@9codestudio.com & 1/1 (100.0\%) \\
\hline
CNT Interaktif Bilgi Tek. Yaz. San. ve Tic. A.S. & 1/1 (100.0\%) \\
\hline
EZTech Apps & 1/1 (100.0\%) \\
\hline
QR SCAN Team & 1/1 (100.0\%) \\
\hline
Security Lab. & 1/1 (100.0\%) \\
\hline
Ultimate Guitar USA LLC & 1/1 (100.0\%) \\
\hline
Radio FM Online Free & 1/1 (100.0\%) \\
\hline
LUCKY YOUTH AND FAMILY SERVICES INC & 1/1 (100.0\%) \\
\hline
RadioFM & 1/1 (100.0\%) \\
\hline
tobbychan21@gmail.com & 1/1 (100.0\%) \\
\hline
Appgeneration - Radio, Podcasts, Games & 1/1 (100.0\%) \\
\hline
Letterboxd Limited & 1/1 (100.0\%) \\
\hline
MeetMe.com & 1/1 (100.0\%) \\
\hline
Skywork AI Pte. Ltd. & 1/1 (100.0\%) \\
\hline
Gamehaus Network & 1/1 (100.0\%) \\
\hline
JoyMore GAME & 1/1 (100.0\%) \\
\hline
BrainMount Ltd & 1/1 (100.0\%) \\
\hline
LeeNgooc & 1/1 (100.0\%) \\
\hline
Poki & 1/1 (100.0\%) \\
\hline
Boom Studio Limited & 1/1 (100.0\%) \\
\hline
BattleCry HQ Studio & 1/1 (100.0\%) \\
\hline
Cricbuzz.com & 1/1 (100.0\%) \\
\hline
Gluak srl & 1/1 (100.0\%) \\
\hline
HealthTracker Apps & 1/1 (100.0\%) \\
\hline
\end{tabular}
\end{table}

\begin{table}[t]
\centering
\begin{tabular}{|p{0.68\linewidth}|p{0.22\linewidth}|}
\hline
Lite Tools Games & 1/1 (100.0\%) \\
\hline
PlayStudioInc & 1/1 (100.0\%) \\
\hline
ZeptoLab & 1/1 (100.0\%) \\
\hline
Smart Widget Labs Co Ltd & 1/1 (100.0\%) \\
\hline
Severex & 1/1 (100.0\%) \\
\hline
LinkDesks - Jewel Games Star & 1/1 (100.0\%) \\
\hline
supermt & 1/1 (100.0\%) \\
\hline
Cricim world & 1/1 (100.0\%) \\
\hline
MEDIADECODE & 1/1 (100.0\%) \\
\hline
Peaksel Ringtones Apps & 1/1 (100.0\%) \\
\hline
Goldlab Pro & 1/1 (100.0\%) \\
\hline
PlayCreek Games & 1/1 (100.0\%) \\
\hline
TAKBIR & 1/1 (100.0\%) \\
\hline
Blackout Lab & 1/1 (100.0\%) \\
\hline
CloudWest Technology & 1/1 (100.0\%) \\
\hline
Touchzing Media Private Limited & 1/1 (100.0\%) \\
\hline
neilbenecke529@gmail.com & 1/1 (100.0\%) \\
\hline
Smart Utils Dev Team & 1/1 (100.0\%) \\
\hline
ruivop & 1/1 (100.0\%) \\
\hline
Wavez Technology Ltd & 1/1 (100.0\%) \\
\hline
Simple Design Ltd. & 1/1 (100.0\%) \\
\hline
AI Screen Translator & 1/1 (100.0\%) \\
\hline
Sad Panda Studios Ltd & 1/1 (100.0\%) \\
\hline
Slimmerbits LLC & 1/1 (100.0\%) \\
\hline
Bright Prospect & 1/1 (100.0\%) \\
\hline
Rear Window Limited & 1/1 (100.0\%) \\
\hline
Appcentric Team & 1/1 (100.0\%) \\
\hline
Trusted Android App & 1/1 (100.0\%) \\
\hline
blurbackgroundstudio@gmail.com & 1/1 (100.0\%) \\
\hline
Apfolife & 1/1 (100.0\%) \\
\hline
abdallahshure99@gmail.com & 1/1 (100.0\%) \\
\hline
musicechofeedback@outlook.com & 1/1 (100.0\%) \\
\hline
Naz Digital & 1/1 (100.0\%) \\
\hline
tuyetnhungpham2710@gmail.com & 1/1 (100.0\%) \\
\hline
TEEWEE LIMITED & 1/1 (100.0\%) \\
\hline
Tiki Punch Studio & 1/1 (100.0\%) \\
\hline
MBit Music Inc. & 1/1 (100.0\%) \\
\hline
Zayzik : LED Keyboard Studio & 1/1 (100.0\%) \\
\hline
Live Wallpapers by Wave Studio & 1/1 (100.0\%) \\
\hline
Digital Dreamworks Studio & 1/1 (100.0\%) \\
\hline
Color Joy & 1/1 (100.0\%) \\
\hline
Weather Radar Team & 1/1 (100.0\%) \\
\hline
Maxlabs Graphic Design Tools & 1/1 (100.0\%) \\
\hline
kahlochSto & 1/1 (100.0\%) \\
\hline
GreenTTeaGame & 1/1 (100.0\%) \\
\hline
GuonianGame & 1/1 (100.0\%) \\
\hline
CHAPTER 4 & 1/1 (100.0\%) \\
\hline
Playa Games & 1/1 (100.0\%) \\
\hline
3D Viet Ha & 1/1 (100.0\%) \\
\hline
Bilge Bulut Mobile & 1/1 (100.0\%) \\
\hline
\end{tabular}
\end{table}

\begin{table}[t]
\centering
\begin{tabular}{|p{0.68\linewidth}|p{0.22\linewidth}|}
\hline
profoundboradwang@gmail.com & 1/1 (100.0\%) \\
\hline
Dream Space & 1/1 (100.0\%) \\
\hline
Ringtone Phone App & 1/1 (100.0\%) \\
\hline
blomsterstudio1733@gmail.com & 1/1 (100.0\%) \\
\hline
BG.Studio & 1/1 (100.0\%) \\
\hline
GREEVIL'S GREED & 1/1 (100.0\%) \\
\hline
Wego.com & 1/1 (100.0\%) \\
\hline
Hitapps Games & 1/1 (100.0\%) \\
\hline
AppOn Innovate & 1/1 (100.0\%) \\
\hline
SZYJ Technology & 1/1 (100.0\%) \\
\hline
AVIRISE LIMITED HK & 1/1 (100.0\%) \\
\hline
Hydodo & 1/1 (100.0\%) \\
\hline
YoYo Dress Up Games & 1/1 (100.0\%) \\
\hline
Addons and Mods for Minecraft & 1/1 (100.0\%) \\
\hline
WE CENTER & 1/1 (100.0\%) \\
\hline
Nguyen Công Phuong & 1/1 (100.0\%) \\
\hline
JoyArk Official-Cloud Games & 1/1 (100.0\%) \\
\hline
rosytales & 1/1 (100.0\%) \\
\hline
dreamphotolab2016@gmail.com & 1/1 (100.0\%) \\
\hline
Studio Sol Comunicação Digital & 1/1 (100.0\%) \\
\hline
Dream Tools & 1/1 (100.0\%) \\
\hline
nttc.studio@gmail.com & 1/1 (100.0\%) \\
\hline
haiyanstore & 1/1 (100.0\%) \\
\hline
LANGUAGE POWER & 1/1 (100.0\%) \\
\hline
1MB Apps Studio & 1/1 (100.0\%) \\
\hline
GOMIN MOBILE & 1/1 (100.0\%) \\
\hline
Beloud.com & 1/1 (100.0\%) \\
\hline
Opera & 1/1 (100.0\%) \\
\hline
Tradera & 1/1 (100.0\%) \\
\hline
NIGHP SOFTWARE & 1/1 (100.0\%) \\
\hline
Golden Gate Media & 1/1 (100.0\%) \\
\hline
No dev found & 16/24 (66.67\%) \\
\hline
Vidmark Inc. & 2/3 (66.67\%) \\
\hline
GeniusTools Labs & 2/3 (66.67\%) \\
\hline
Cards & 3/5 (60.0\%) \\
\hline
Galaxy studio apps & 2/4 (50.0\%) \\
\hline
CEM SOFTWARE LTD & 2/4 (50.0\%) \\
\hline
Gamma Play & 1/2 (50.0\%) \\
\hline
Vidow™ & 1/2 (50.0\%) \\
\hline
Fillog Studio & 1/2 (50.0\%) \\
\hline
gobistudio88@gmail.com & 1/2 (50.0\%) \\
\hline
DogByte Games & 1/3 (33.33\%) \\
\hline
XGAME STUDIO & 1/3 (33.33\%) \\
\hline
Data Unavailable & 0/11 (0\%) \\
\hline
KIGLE & 0/8 (0\%) \\
\hline
GunjanApps Studios & 0/6 (0\%) \\
\hline
One Music Player & 0/4 (0\%) \\
\hline
App Game Development Solutions & 0/3 (0\%) \\
\hline
Shadow soft & 0/3 (0\%) \\
\hline
Baram FZE & 0/3 (0\%) \\
\hline
Dumitru Boico & 0/3 (0\%) \\
\hline
\end{tabular}
\end{table}

\begin{table}[t]
\centering
\begin{tabular}{|p{0.68\linewidth}|p{0.22\linewidth}|}
\hline
P \& L Studio & 0/3 (0\%) \\
\hline
ElePant: Kids Learning Games for Toddlers \& Baby & 0/3 (0\%) \\
\hline
Three Cookers Game & 0/3 (0\%) \\
\hline
Awesome Game Studio & 0/2 (0\%) \\
\hline
WallForApps & 0/2 (0\%) \\
\hline
TarrySoft & 0/2 (0\%) \\
\hline
IDZ Digital Private Limited & 0/2 (0\%) \\
\hline
InShot Video Editor & 0/2 (0\%) \\
\hline
Anfona Tech & 0/2 (0\%) \\
\hline
MayZing Tech & 0/2 (0\%) \\
\hline
FastSoft & 0/2 (0\%) \\
\hline
Flavapp & 0/2 (0\%) \\
\hline
Masivapp & 0/2 (0\%) \\
\hline
Stillfront Supremacy GmbH & 0/2 (0\%) \\
\hline
Yangmei Studios & 0/2 (0\%) \\
\hline
Banix Studio & 0/2 (0\%) \\
\hline
DOSA Apps & 0/2 (0\%) \\
\hline
Firehawk & 0/2 (0\%) \\
\hline
Weather Forecast \& Widget \& Radar & 0/2 (0\%) \\
\hline
Amila & 0/2 (0\%) \\
\hline
dev@pocketclubs.com & 0/2 (0\%) \\
\hline
Fabulous Fun & 0/1 (0\%) \\
\hline
Panteon & 0/1 (0\%) \\
\hline
Catchy Tools & 0/1 (0\%) \\
\hline
M2Catalyst, LLC. & 0/1 (0\%) \\
\hline
trantheanhmkt@gmail.com & 0/1 (0\%) \\
\hline
Kongs & 0/1 (0\%) \\
\hline
Amobear Studio & 0/1 (0\%) \\
\hline
Binary Cores & 0/1 (0\%) \\
\hline
Atlantis Ultra Station & 0/1 (0\%) \\
\hline
Launchers World & 0/1 (0\%) \\
\hline
recorder \& smart apps & 0/1 (0\%) \\
\hline
CoolDev Co Ltd & 0/1 (0\%) \\
\hline
VIDEOSHOW Video Editor \& Maker \& AI Chat Generator & 0/1 (0\%) \\
\hline
Video \& Photo Editor Apps & 0/1 (0\%) \\
\hline
Fast Tour Booking - Flights \& Hotels & 0/1 (0\%) \\
\hline
100Pi Labs & 0/1 (0\%) \\
\hline
hello@yangmeistudios.com & 0/1 (0\%) \\
\hline
Yanstar Studio OU & 0/1 (0\%) \\
\hline
Generation z apps & 0/1 (0\%) \\
\hline
Apps You Love & 0/1 (0\%) \\
\hline
Arioch Pds - Apps \& Games & 0/1 (0\%) \\
\hline
Muslim Worldapp & 0/1 (0\%) \\
\hline
Jolt Global Apps - VPN, Ai, Keyboard \& Learning & 0/1 (0\%) \\
\hline
Mabixa & 0/1 (0\%) \\
\hline
GeDa DevTeam & 0/1 (0\%) \\
\hline
SincMobile Apps & 0/1 (0\%) \\
\hline
renvosoft@gmail.com & 0/1 (0\%) \\
\hline
anhtuannt3011@gmail.com & 0/1 (0\%) \\
\hline
Appsomniacs LLC & 0/1 (0\%) \\
\hline
\end{tabular}
\end{table}

\begin{table}[t]
\centering
\begin{tabular}{|p{0.68\linewidth}|p{0.22\linewidth}|}
\hline
SOUSSI NIAIMI Badr-Eddine & 0/1 (0\%) \\
\hline
Mediocre & 0/1 (0\%) \\
\hline
HAU SALA GAME COMPANY LIMITED & 0/1 (0\%) \\
\hline
Flirtify: Live Chat \& Dating & 0/1 (0\%) \\
\hline
Stillfront Supremacy Ltd & 0/1 (0\%) \\
\hline
Apps Resort - Daily Tool Apps & 0/1 (0\%) \\
\hline
phamnguyetnt89@gmail.com & 0/1 (0\%) \\
\hline
Love Photo Frames & 0/1 (0\%) \\
\hline
Fitify Workouts s.r.o. & 0/1 (0\%) \\
\hline
Litera Games & 0/1 (0\%) \\
\hline
Okapps & 0/1 (0\%) \\
\hline
Pixel Kraft Studios & 0/1 (0\%) \\
\hline
Wow Themes \& Fulll HD Themes & 0/1 (0\%) \\
\hline
Gigo Ltc & 0/1 (0\%) \\
\hline
TM company & 0/1 (0\%) \\
\hline
tnifoui wallpaper & 0/1 (0\%) \\
\hline
Appache apps and games ltd & 0/1 (0\%) \\
\hline
ng-labs & 0/1 (0\%) \\
\hline
Freepik Company & 0/1 (0\%) \\
\hline
ChargeFinder & 0/1 (0\%) \\
\hline
Quality App Zone & 0/1 (0\%) \\
\hline
BoostVision & 0/1 (0\%) \\
\hline
StarPham & 0/1 (0\%) \\
\hline
text messages & 0/1 (0\%) \\
\hline
Bringar Apps & 0/1 (0\%) \\
\hline
Quarzo Apps & 0/1 (0\%) \\
\hline
JMGame & 0/1 (0\%) \\
\hline
Travel Maps Tech & 0/1 (0\%) \\
\hline
Copa Fácil & 0/1 (0\%) \\
\hline
Messier31 & 0/1 (0\%) \\
\hline
Accurate Weather Forecast \& Weather Radar Map & 0/1 (0\%) \\
\hline
TOH Talent Team & 0/1 (0\%) \\
\hline
Block Puzzle Games 2018 & 0/1 (0\%) \\
\hline
EriitoDraw & 0/1 (0\%) \\
\hline
Music Apps - Allmusic & 0/1 (0\%) \\
\hline
nguyenquynhtrangkt08@gmail.com & 0/1 (0\%) \\
\hline
neraldoverland342@gmail.com & 0/1 (0\%) \\
\hline
BoBo World Games & 0/1 (0\%) \\
\hline
probadoSoft & 0/1 (0\%) \\
\hline
TOH Games & 0/1 (0\%) \\
\hline
Minibuu & 0/1 (0\%) \\
\hline
Tool Apps Hub & 0/1 (0\%) \\
\hline
App Bards & 0/1 (0\%) \\
\hline
Digital App Valley & 0/1 (0\%) \\
\hline
SoftAxes & 0/1 (0\%) \\
\hline
GjangHa & 0/1 (0\%) \\
\hline
IApplication & 0/1 (0\%) \\
\hline
Konnect Apps & 0/1 (0\%) \\
\hline
abdulmuttalip.er4680@gmail.com & 0/1 (0\%) \\
\hline
Marcel Bartecki & 0/1 (0\%) \\
\hline
\end{tabular}
\end{table}

\begin{table}[t]
\centering
\begin{tabular}{|p{0.68\linewidth}|p{0.22\linewidth}|}
\hline
Splaish Studio & 0/1 (0\%) \\
\hline
Kolmo Games & 0/1 (0\%) \\
\hline
Plus AI & 0/1 (0\%) \\
\hline
QR Code Scanner \& Barcode Reader & 0/1 (0\%) \\
\hline
Kamrej Apps & 0/1 (0\%) \\
\hline
Nova Apps Studios & 0/1 (0\%) \\
\hline
BigSoft inc. & 0/1 (0\%) \\
\hline
Sun global & 0/1 (0\%) \\
\hline
Easy Language Translator & 0/1 (0\%) \\
\hline
nhuomtv@cemsoftwareltd.com & 0/1 (0\%) \\
\hline
mobirix & 0/1 (0\%) \\
\hline
Lincod Studio & 0/1 (0\%) \\
\hline
Fit Health Inc. & 0/1 (0\%) \\
\hline
Tulip Sports TV & 0/1 (0\%) \\
\hline
Dictionary World11 & 0/1 (0\%) \\
\hline
Appslogie & 0/1 (0\%) \\
\hline
appwave333@gmail.com & 0/1 (0\%) \\
\hline
genie islam & 0/1 (0\%) \\
\hline
PhucArts & 0/1 (0\%) \\
\hline
Mi Game Pro Uru & 0/1 (0\%) \\
\hline
Rubén Mayayo & 0/1 (0\%) \\
\hline
Tools Generation Hub & 0/1 (0\%) \\
\hline
tracyhoggappstore@gmail.com & 0/1 (0\%) \\
\hline
Radiant Islamic Apps & 0/1 (0\%) \\
\hline
UniTiki & 0/1 (0\%) \\
\hline
InPics & 0/1 (0\%) \\
\hline
highsecure & 0/1 (0\%) \\
\hline
QY Studio & 0/1 (0\%) \\
\hline
Greenstream Apps & 0/1 (0\%) \\
\hline
Simple Mobile Tool & 0/1 (0\%) \\
\hline
Viaplay & 0/1 (0\%) \\
\hline
\end{tabular} 
\vspace{5mm}
\end{table}

\begin{table}[t]
\caption{Developer active stage data}
\label{tab:violating_developers_stage_2}

\centering
\begin{tabular}{|p{0.68\linewidth}|p{0.22\linewidth}|}
\hline
\textbf{Developer (name/email)} & \textbf{Violation \%} \\
\hline
Easybrain & 14/14 (100.0\%) \\
\hline
BoomBit Games & 7/7 (100.0\%) \\
\hline
CEM SOFTWARE LTD & 4/4 (100.0\%) \\
\hline
RobTop Games & 4/4 (100.0\%) \\
\hline
nguyenluan@lutech.ltd & 4/4 (100.0\%) \\
\hline
hoahuongduong170120@gmail.com & 4/4 (100.0\%) \\
\hline
SayGames Ltd & 3/3 (100.0\%) \\
\hline
Prometheus Interactive LLC & 3/3 (100.0\%) \\
\hline
LoveColoring Game & 3/3 (100.0\%) \\
\hline
KAYAC Inc. & 3/3 (100.0\%) \\
\hline
ZenLife Games Ltd & 3/3 (100.0\%) \\
\hline
ThemeKit & 3/3 (100.0\%) \\
\hline
XGAME STUDIO & 3/3 (100.0\%) \\
\hline
IVYGAMES & 3/3 (100.0\%) \\
\hline
FMPROJECT LIMITED & 2/2 (100.0\%) \\
\hline
LifeSim & 2/2 (100.0\%) \\
\hline
KamaGames & 2/2 (100.0\%) \\
\hline
publishing.apero@gmail.com & 2/2 (100.0\%) \\
\hline
LifePulse Puzzle Game Studio & 2/2 (100.0\%) \\
\hline
Gameberry Labs & 2/2 (100.0\%) \\
\hline
GAME OFFLINE HAY & 2/2 (100.0\%) \\
\hline
Hippo Lab & 2/2 (100.0\%) \\
\hline
MAGIC SEVEN CO., LIMITED & 2/2 (100.0\%) \\
\hline
HDuo Fun Games & 2/2 (100.0\%) \\
\hline
Maxlabs Photo Editor & 2/2 (100.0\%) \\
\hline
Color App Team & 2/2 (100.0\%) \\
\hline
FunGear inc & 2/2 (100.0\%) \\
\hline
Eyewind & 2/2 (100.0\%) \\
\hline
Braly JSC & 2/2 (100.0\%) \\
\hline
vinhdn07@gmail.com & 2/2 (100.0\%) \\
\hline
admin@vtn.global & 2/2 (100.0\%) \\
\hline
SUPERIOR  STUDIO & 2/2 (100.0\%) \\
\hline
Appwest Limited & 2/2 (100.0\%) \\
\hline
Metaverse Labs & 2/2 (100.0\%) \\
\hline
Beat Blend Labs & 2/2 (100.0\%) \\
\hline
Red Sky Labs & 2/2 (100.0\%) \\
\hline
Solitaire Aquarium & 2/2 (100.0\%) \\
\hline
Weather Forecast \& Widget \& Radar & 2/2 (100.0\%) \\
\hline
Fillog Studio & 2/2 (100.0\%) \\
\hline
Mortys Games & 2/2 (100.0\%) \\
\hline
gobistudio88@gmail.com & 2/2 (100.0\%) \\
\hline
Adkins Studio & 2/2 (100.0\%) \\
\hline
Dainik Bhaskar Group & 2/2 (100.0\%) \\
\hline
DIC-o & 1/1 (100.0\%) \\
\hline
mobileapps@tubema.ltd & 1/1 (100.0\%) \\
\hline
Rainberry, Inc. & 1/1 (100.0\%) \\
\hline
ORANGE GAME & 1/1 (100.0\%) \\
\hline
Fearless Games sp. j. & 1/1 (100.0\%) \\
\hline
Escape Adventure Games & 1/1 (100.0\%) \\
\hline
WONDER GROUP & 1/1 (100.0\%) \\
\hline
\end{tabular}
\end{table}

\begin{table}[t]
\centering
\begin{tabular}{|p{0.68\linewidth}|p{0.22\linewidth}|}
\hline
Estoty & 1/1 (100.0\%) \\
\hline
Supercent, Inc. & 1/1 (100.0\%) \\
\hline
cerdillac & 1/1 (100.0\%) \\
\hline
mulanidhvani@gmail.com & 1/1 (100.0\%) \\
\hline
FALCON GAME & 1/1 (100.0\%) \\
\hline
trantheanhmkt@gmail.com & 1/1 (100.0\%) \\
\hline
TFive  Labs & 1/1 (100.0\%) \\
\hline
changpeng & 1/1 (100.0\%) \\
\hline
Amobear Studio & 1/1 (100.0\%) \\
\hline
Dream Tap & 1/1 (100.0\%) \\
\hline
OUTLOU:D GAMES & 1/1 (100.0\%) \\
\hline
Better World Games & 1/1 (100.0\%) \\
\hline
Nebuchadnezzar DOO & 1/1 (100.0\%) \\
\hline
KTW Apps & 1/1 (100.0\%) \\
\hline
Free VPN Planet & 1/1 (100.0\%) \\
\hline
Digipom & 1/1 (100.0\%) \\
\hline
recorder \& smart apps & 1/1 (100.0\%) \\
\hline
Audify Player. & 1/1 (100.0\%) \\
\hline
Colorful Point & 1/1 (100.0\%) \\
\hline
dammuhzaj@gmail.com & 1/1 (100.0\%) \\
\hline
Diavostar PTE. LTD & 1/1 (100.0\%) \\
\hline
CoolDev Co Ltd & 1/1 (100.0\%) \\
\hline
VIDEOSHOW Video Editor \& Maker \& AI Chat Generator & 1/1 (100.0\%) \\
\hline
developer@vottak.app & 1/1 (100.0\%) \\
\hline
Lotum one GmbH & 1/1 (100.0\%) \\
\hline
Clappy LLC & 1/1 (100.0\%) \\
\hline
Andrey Solovyev & 1/1 (100.0\%) \\
\hline
Qbis Studio & 1/1 (100.0\%) \\
\hline
Loop Games & 1/1 (100.0\%) \\
\hline
ZenTint Creations & 1/1 (100.0\%) \\
\hline
7788`s & 1/1 (100.0\%) \\
\hline
MOUNTAIN GAME & 1/1 (100.0\%) \\
\hline
Doodle Mobile Ltd. & 1/1 (100.0\%) \\
\hline
CanaryDroid & 1/1 (100.0\%) \\
\hline
Macro Tap & 1/1 (100.0\%) \\
\hline
1SOFT & 1/1 (100.0\%) \\
\hline
renvosoft@gmail.com & 1/1 (100.0\%) \\
\hline
Emily Yu & 1/1 (100.0\%) \\
\hline
Prop studio & 1/1 (100.0\%) \\
\hline
haphuong4251@gmail.com & 1/1 (100.0\%) \\
\hline
Fancolor & 1/1 (100.0\%) \\
\hline
Miniclip.com & 1/1 (100.0\%) \\
\hline
CASUAL AZUR GAMES & 1/1 (100.0\%) \\
\hline
Theago Liddell & 1/1 (100.0\%) \\
\hline
Warrior Game & 1/1 (100.0\%) \\
\hline
Pixel Art - draw in fun & 1/1 (100.0\%) \\
\hline
AppQuantum & 1/1 (100.0\%) \\
\hline
EasyFun Puzzle Game Studio & 1/1 (100.0\%) \\
\hline
ZiMAD & 1/1 (100.0\%) \\
\hline
RED BRIX COMPUTER SYSTEMS & 1/1 (100.0\%) \\
\hline
\end{tabular}
\end{table}

\begin{table}[t]
\centering
\begin{tabular}{|p{0.68\linewidth}|p{0.22\linewidth}|}
\hline
Focus apps & 1/1 (100.0\%) \\
\hline
Kitten Doll HK & 1/1 (100.0\%) \\
\hline
Kitten doll & 1/1 (100.0\%) \\
\hline
Flirtify: Live Chat \& Dating & 1/1 (100.0\%) \\
\hline
Meeto Team & 1/1 (100.0\%) \\
\hline
BUZZMEDIA INC. & 1/1 (100.0\%) \\
\hline
Apps Resort - Daily Tool Apps & 1/1 (100.0\%) \\
\hline
Smart AI DEV & 1/1 (100.0\%) \\
\hline
checksmartapp@gmail.com & 1/1 (100.0\%) \\
\hline
cghxstudio@gmail.com & 1/1 (100.0\%) \\
\hline
aimirror-support@polyversestudio.com & 1/1 (100.0\%) \\
\hline
Outdoing Apps & 1/1 (100.0\%) \\
\hline
Pixl Concerto Tech & 1/1 (100.0\%) \\
\hline
o16i Apps & 1/1 (100.0\%) \\
\hline
Clever Apps Pte. Ltd. & 1/1 (100.0\%) \\
\hline
Coreup Teknoloji Limited Şirketi & 1/1 (100.0\%) \\
\hline
Funny Design Keyboard Themes & 1/1 (100.0\%) \\
\hline
Delicate theme for Android App & 1/1 (100.0\%) \\
\hline
Wow Themes \& Fulll HD Themes & 1/1 (100.0\%) \\
\hline
Esame Marketing Limited & 1/1 (100.0\%) \\
\hline
Fancy Studio Mods & 1/1 (100.0\%) \\
\hline
Live Wallpapers and Emoji Keyboard Themes & 1/1 (100.0\%) \\
\hline
chudang@9codestudio.com & 1/1 (100.0\%) \\
\hline
CNT Interaktif Bilgi Tek. Yaz. San. ve Tic. A.S. & 1/1 (100.0\%) \\
\hline
EZTech Apps & 1/1 (100.0\%) \\
\hline
QR SCAN Team & 1/1 (100.0\%) \\
\hline
Security Lab. & 1/1 (100.0\%) \\
\hline
Ultimate Guitar USA LLC & 1/1 (100.0\%) \\
\hline
Radio FM Online Free & 1/1 (100.0\%) \\
\hline
RadioFM & 1/1 (100.0\%) \\
\hline
tobbychan21@gmail.com & 1/1 (100.0\%) \\
\hline
Appgeneration - Radio, Podcasts, Games & 1/1 (100.0\%) \\
\hline
MeetMe.com & 1/1 (100.0\%) \\
\hline
Skywork AI Pte. Ltd. & 1/1 (100.0\%) \\
\hline
Gamehaus Network & 1/1 (100.0\%) \\
\hline
JoyMore GAME & 1/1 (100.0\%) \\
\hline
BrainMount Ltd & 1/1 (100.0\%) \\
\hline
LeeNgooc & 1/1 (100.0\%) \\
\hline
Poki & 1/1 (100.0\%) \\
\hline
Boom Studio Limited & 1/1 (100.0\%) \\
\hline
JMGame & 1/1 (100.0\%) \\
\hline
BattleCry HQ Studio & 1/1 (100.0\%) \\
\hline
Cricbuzz.com & 1/1 (100.0\%) \\
\hline
Messier31 & 1/1 (100.0\%) \\
\hline
Gluak srl & 1/1 (100.0\%) \\
\hline
HealthTracker Apps & 1/1 (100.0\%) \\
\hline
Lite Tools Games & 1/1 (100.0\%) \\
\hline
PlayStudioInc & 1/1 (100.0\%) \\
\hline
ZeptoLab & 1/1 (100.0\%) \\
\hline
Smart Widget Labs Co Ltd & 1/1 (100.0\%) \\
\hline
\end{tabular}
\end{table}

\begin{table}[t]
\centering
\begin{tabular}{|p{0.68\linewidth}|p{0.22\linewidth}|}
\hline
LinkDesks - Jewel Games Star & 1/1 (100.0\%) \\
\hline
supermt & 1/1 (100.0\%) \\
\hline
Cricim world & 1/1 (100.0\%) \\
\hline
MEDIADECODE & 1/1 (100.0\%) \\
\hline
Peaksel Ringtones Apps & 1/1 (100.0\%) \\
\hline
PlayCreek Games & 1/1 (100.0\%) \\
\hline
CloudWest Technology & 1/1 (100.0\%) \\
\hline
Touchzing Media Private Limited & 1/1 (100.0\%) \\
\hline
neilbenecke529@gmail.com & 1/1 (100.0\%) \\
\hline
ruivop & 1/1 (100.0\%) \\
\hline
Wavez Technology Ltd & 1/1 (100.0\%) \\
\hline
Tool Apps Hub & 1/1 (100.0\%) \\
\hline
AI Screen Translator & 1/1 (100.0\%) \\
\hline
Slimmerbits LLC & 1/1 (100.0\%) \\
\hline
Bright Prospect & 1/1 (100.0\%) \\
\hline
Rear Window Limited & 1/1 (100.0\%) \\
\hline
Appcentric Team & 1/1 (100.0\%) \\
\hline
Trusted Android App & 1/1 (100.0\%) \\
\hline
blurbackgroundstudio@gmail.com & 1/1 (100.0\%) \\
\hline
Apfolife & 1/1 (100.0\%) \\
\hline
abdallahshure99@gmail.com & 1/1 (100.0\%) \\
\hline
musicechofeedback@outlook.com & 1/1 (100.0\%) \\
\hline
GjangHa & 1/1 (100.0\%) \\
\hline
Naz Digital & 1/1 (100.0\%) \\
\hline
Konnect Apps & 1/1 (100.0\%) \\
\hline
tuyetnhungpham2710@gmail.com & 1/1 (100.0\%) \\
\hline
TEEWEE LIMITED & 1/1 (100.0\%) \\
\hline
QR Code Scanner \& Barcode Reader & 1/1 (100.0\%) \\
\hline
Tiki Punch Studio & 1/1 (100.0\%) \\
\hline
MBit Music Inc. & 1/1 (100.0\%) \\
\hline
Zayzik : LED Keyboard Studio & 1/1 (100.0\%) \\
\hline
Live Wallpapers by Wave Studio & 1/1 (100.0\%) \\
\hline
Digital Dreamworks Studio & 1/1 (100.0\%) \\
\hline
Color Joy & 1/1 (100.0\%) \\
\hline
Weather Radar Team & 1/1 (100.0\%) \\
\hline
Maxlabs Graphic Design Tools & 1/1 (100.0\%) \\
\hline
kahlochSto & 1/1 (100.0\%) \\
\hline
nhuomtv@cemsoftwareltd.com & 1/1 (100.0\%) \\
\hline
GreenTTeaGame & 1/1 (100.0\%) \\
\hline
GuonianGame & 1/1 (100.0\%) \\
\hline
3D Viet Ha & 1/1 (100.0\%) \\
\hline
Bilge Bulut Mobile & 1/1 (100.0\%) \\
\hline
profoundboradwang@gmail.com & 1/1 (100.0\%) \\
\hline
Dream Space & 1/1 (100.0\%) \\
\hline
Ringtone Phone App & 1/1 (100.0\%) \\
\hline
blomsterstudio1733@gmail.com & 1/1 (100.0\%) \\
\hline
BG.Studio & 1/1 (100.0\%) \\
\hline
Wego.com & 1/1 (100.0\%) \\
\hline
Hitapps Games & 1/1 (100.0\%) \\
\hline
Fit Health Inc. & 1/1 (100.0\%) \\
\hline
\end{tabular}
\end{table}

\begin{table}[t]
\centering
\begin{tabular}{|p{0.68\linewidth}|p{0.22\linewidth}|}
\hline
AppOn Innovate & 1/1 (100.0\%) \\
\hline
SZYJ Technology & 1/1 (100.0\%) \\
\hline
AVIRISE LIMITED HK & 1/1 (100.0\%) \\
\hline
Hydodo & 1/1 (100.0\%) \\
\hline
YoYo Dress Up Games & 1/1 (100.0\%) \\
\hline
Addons and Mods for Minecraft & 1/1 (100.0\%) \\
\hline
WE CENTER & 1/1 (100.0\%) \\
\hline
Nguyen Công Phuong & 1/1 (100.0\%) \\
\hline
JoyArk Official-Cloud Games & 1/1 (100.0\%) \\
\hline
rosytales & 1/1 (100.0\%) \\
\hline
dreamphotolab2016@gmail.com & 1/1 (100.0\%) \\
\hline
Tools Generation Hub & 1/1 (100.0\%) \\
\hline
Dream Tools & 1/1 (100.0\%) \\
\hline
nttc.studio@gmail.com & 1/1 (100.0\%) \\
\hline
LANGUAGE POWER & 1/1 (100.0\%) \\
\hline
1MB Apps Studio & 1/1 (100.0\%) \\
\hline
QY Studio & 1/1 (100.0\%) \\
\hline
Beloud.com & 1/1 (100.0\%) \\
\hline
Opera & 1/1 (100.0\%) \\
\hline
Tradera & 1/1 (100.0\%) \\
\hline
Golden Gate Media & 1/1 (100.0\%) \\
\hline
Outfit7 Limited & 7/9 (77.78\%) \\
\hline
gameone & 3/4 (75.0\%) \\
\hline
Vidmark Inc. & 2/3 (66.67\%) \\
\hline
P \& L Studio & 2/3 (66.67\%) \\
\hline
GeniusTools Labs & 2/3 (66.67\%) \\
\hline
trusted.mobile.app@gmail.com & 2/3 (66.67\%) \\
\hline
Cards & 3/5 (60.0\%) \\
\hline
Battery Stats Saver & 3/5 (60.0\%) \\
\hline
No dev found & 13/24 (54.17\%) \\
\hline
Galaxy studio apps & 2/4 (50.0\%) \\
\hline
Gamma Play & 1/2 (50.0\%) \\
\hline
Vidow™ & 1/2 (50.0\%) \\
\hline
DogByte Games & 1/3 (33.33\%) \\
\hline
Three Cookers Game & 1/3 (33.33\%) \\
\hline
Data Unavailable & 0/11 (0\%) \\
\hline
KIGLE & 0/8 (0\%) \\
\hline
GunjanApps Studios & 0/6 (0\%) \\
\hline
CrazyLabs LTD & 0/5 (0\%) \\
\hline
One Music Player & 0/4 (0\%) \\
\hline
App Game Development Solutions & 0/3 (0\%) \\
\hline
Shadow soft & 0/3 (0\%) \\
\hline
Baram FZE & 0/3 (0\%) \\
\hline
Coco Play By TabTale & 0/3 (0\%) \\
\hline
Dumitru Boico & 0/3 (0\%) \\
\hline
ElePant: Kids Learning Games for Toddlers \& Baby & 0/3 (0\%) \\
\hline
Awesome Game Studio & 0/2 (0\%) \\
\hline
WallForApps & 0/2 (0\%) \\
\hline
TarrySoft & 0/2 (0\%) \\
\hline
IDZ Digital Private Limited & 0/2 (0\%) \\
\hline
InShot Video Editor & 0/2 (0\%) \\
\hline
\end{tabular}
\end{table}

\begin{table}[t]
\centering
\begin{tabular}{|p{0.68\linewidth}|p{0.22\linewidth}|}
\hline
Anfona Tech & 0/2 (0\%) \\
\hline
MayZing Tech & 0/2 (0\%) \\
\hline
FastSoft & 0/2 (0\%) \\
\hline
Flavapp & 0/2 (0\%) \\
\hline
Masivapp & 0/2 (0\%) \\
\hline
Stillfront Supremacy GmbH & 0/2 (0\%) \\
\hline
Yangmei Studios & 0/2 (0\%) \\
\hline
Banix Studio & 0/2 (0\%) \\
\hline
DOSA Apps & 0/2 (0\%) \\
\hline
Firehawk & 0/2 (0\%) \\
\hline
Amila & 0/2 (0\%) \\
\hline
HK Hero Entertainment Co., Limited & 0/2 (0\%) \\
\hline
dev@pocketclubs.com & 0/2 (0\%) \\
\hline
Gianluca Cisana & 0/1 (0\%) \\
\hline
TabTale & 0/1 (0\%) \\
\hline
Fabulous Fun & 0/1 (0\%) \\
\hline
Panteon & 0/1 (0\%) \\
\hline
Catchy Tools & 0/1 (0\%) \\
\hline
M2Catalyst, LLC. & 0/1 (0\%) \\
\hline
Kongs & 0/1 (0\%) \\
\hline
Binary Cores & 0/1 (0\%) \\
\hline
Atlantis Ultra Station & 0/1 (0\%) \\
\hline
Launchers World & 0/1 (0\%) \\
\hline
Siti Berkah Studio & 0/1 (0\%) \\
\hline
Video \& Photo Editor Apps & 0/1 (0\%) \\
\hline
Fast Tour Booking - Flights \& Hotels & 0/1 (0\%) \\
\hline
ASD Dev Video Player for All Format & 0/1 (0\%) \\
\hline
100Pi Labs & 0/1 (0\%) \\
\hline
Qiiwi Games AB & 0/1 (0\%) \\
\hline
hello@yangmeistudios.com & 0/1 (0\%) \\
\hline
Buddies Games Inc. & 0/1 (0\%) \\
\hline
Yanstar Studio OU & 0/1 (0\%) \\
\hline
Generation z apps & 0/1 (0\%) \\
\hline
Apps You Love & 0/1 (0\%) \\
\hline
Arioch Pds - Apps \& Games & 0/1 (0\%) \\
\hline
Muslim Worldapp & 0/1 (0\%) \\
\hline
Jolt Global Apps - VPN, Ai, Keyboard \& Learning & 0/1 (0\%) \\
\hline
Mabixa & 0/1 (0\%) \\
\hline
GeDa DevTeam & 0/1 (0\%) \\
\hline
SincMobile Apps & 0/1 (0\%) \\
\hline
anhtuannt3011@gmail.com & 0/1 (0\%) \\
\hline
Appsomniacs LLC & 0/1 (0\%) \\
\hline
SOUSSI NIAIMI Badr-Eddine & 0/1 (0\%) \\
\hline
Mediocre & 0/1 (0\%) \\
\hline
HAU SALA GAME COMPANY LIMITED & 0/1 (0\%) \\
\hline
Stillfront Supremacy Ltd & 0/1 (0\%) \\
\hline
phamnguyetnt89@gmail.com & 0/1 (0\%) \\
\hline
sunflower studio & 0/1 (0\%) \\
\hline
Love Photo Frames & 0/1 (0\%) \\
\hline
ITO Technologies, Inc. & 0/1 (0\%) \\
\hline
\end{tabular}
\end{table}

\begin{table}[t]
\centering
\begin{tabular}{|p{0.68\linewidth}|p{0.22\linewidth}|}
\hline
Fitify Workouts s.r.o. & 0/1 (0\%) \\
\hline
Litera Games & 0/1 (0\%) \\
\hline
Scopely & 0/1 (0\%) \\
\hline
Okapps & 0/1 (0\%) \\
\hline
Pixel Kraft Studios & 0/1 (0\%) \\
\hline
uschultz & 0/1 (0\%) \\
\hline
Gigo Ltc & 0/1 (0\%) \\
\hline
App Soft Studio & 0/1 (0\%) \\
\hline
TM company & 0/1 (0\%) \\
\hline
tnifoui wallpaper & 0/1 (0\%) \\
\hline
Appache apps and games ltd & 0/1 (0\%) \\
\hline
ng-labs & 0/1 (0\%) \\
\hline
Freepik Company & 0/1 (0\%) \\
\hline
ChargeFinder & 0/1 (0\%) \\
\hline
Quality App Zone & 0/1 (0\%) \\
\hline
BoostVision & 0/1 (0\%) \\
\hline
LUCKY YOUTH AND FAMILY SERVICES INC & 0/1 (0\%) \\
\hline
Letterboxd Limited & 0/1 (0\%) \\
\hline
StarPham & 0/1 (0\%) \\
\hline
text messages & 0/1 (0\%) \\
\hline
Bringar Apps & 0/1 (0\%) \\
\hline
Quarzo Apps & 0/1 (0\%) \\
\hline
Travel Maps Tech & 0/1 (0\%) \\
\hline
Copa Fácil & 0/1 (0\%) \\
\hline
Accurate Weather Forecast \& Weather Radar Map & 0/1 (0\%) \\
\hline
TOH Talent Team & 0/1 (0\%) \\
\hline
Block Puzzle Games 2018 & 0/1 (0\%) \\
\hline
Severex & 0/1 (0\%) \\
\hline
EriitoDraw & 0/1 (0\%) \\
\hline
Music Apps - Allmusic & 0/1 (0\%) \\
\hline
nguyenquynhtrangkt08@gmail.com & 0/1 (0\%) \\
\hline
Goldlab Pro & 0/1 (0\%) \\
\hline
neraldoverland342@gmail.com & 0/1 (0\%) \\
\hline
BoBo World Games & 0/1 (0\%) \\
\hline
TAKBIR & 0/1 (0\%) \\
\hline
Blackout Lab & 0/1 (0\%) \\
\hline
probadoSoft & 0/1 (0\%) \\
\hline
Smart Utils Dev Team & 0/1 (0\%) \\
\hline
TOH Games & 0/1 (0\%) \\
\hline
Minibuu & 0/1 (0\%) \\
\hline
Simple Design Ltd. & 0/1 (0\%) \\
\hline
Sad Panda Studios Ltd & 0/1 (0\%) \\
\hline
App Bards & 0/1 (0\%) \\
\hline
Digital App Valley & 0/1 (0\%) \\
\hline
SoftAxes & 0/1 (0\%) \\
\hline
IApplication & 0/1 (0\%) \\
\hline
abdulmuttalip.er4680@gmail.com & 0/1 (0\%) \\
\hline
Marcel Bartecki & 0/1 (0\%) \\
\hline
Splaish Studio & 0/1 (0\%) \\
\hline
Kolmo Games & 0/1 (0\%) \\
\hline
\end{tabular}
\end{table}

\begin{table}[t]
\centering
\begin{tabular}{|p{0.68\linewidth}|p{0.22\linewidth}|}
\hline
Plus AI & 0/1 (0\%) \\
\hline
Kamrej Apps & 0/1 (0\%) \\
\hline
Nova Apps Studios & 0/1 (0\%) \\
\hline
BigSoft inc. & 0/1 (0\%) \\
\hline
Sun global & 0/1 (0\%) \\
\hline
Easy Language Translator & 0/1 (0\%) \\
\hline
mobirix & 0/1 (0\%) \\
\hline
CHAPTER 4 & 0/1 (0\%) \\
\hline
Playa Games & 0/1 (0\%) \\
\hline
Lincod Studio & 0/1 (0\%) \\
\hline
GREEVIL'S GREED & 0/1 (0\%) \\
\hline
Tulip Sports TV & 0/1 (0\%) \\
\hline
Dictionary World11 & 0/1 (0\%) \\
\hline
Appslogie & 0/1 (0\%) \\
\hline
appwave333@gmail.com & 0/1 (0\%) \\
\hline
genie islam & 0/1 (0\%) \\
\hline
PhucArts & 0/1 (0\%) \\
\hline
Mi Game Pro Uru & 0/1 (0\%) \\
\hline
Rubén Mayayo & 0/1 (0\%) \\
\hline
Studio Sol Comunicação Digital & 0/1 (0\%) \\
\hline
tracyhoggappstore@gmail.com & 0/1 (0\%) \\
\hline
Radiant Islamic Apps & 0/1 (0\%) \\
\hline
UniTiki & 0/1 (0\%) \\
\hline
InPics & 0/1 (0\%) \\
\hline
haiyanstore & 0/1 (0\%) \\
\hline
highsecure & 0/1 (0\%) \\
\hline
GOMIN MOBILE & 0/1 (0\%) \\
\hline
Greenstream Apps & 0/1 (0\%) \\
\hline
Simple Mobile Tool & 0/1 (0\%) \\
\hline
Viaplay & 0/1 (0\%) \\
\hline
NIGHP SOFTWARE & 0/1 (0\%) \\
\hline
\end{tabular} 
\vspace{5mm}
\end{table}

\end{document}